\documentclass[aip,
 amsmath,amssymb,
 reprint,%
]{revtex4-2}

\usepackage{graphicx}
\usepackage{dcolumn}
\usepackage{bm}

\usepackage[utf8]{inputenc}
\usepackage[T1]{fontenc}
\usepackage{mathptmx}
\usepackage{etoolbox}
\usepackage{hyperref}

\usepackage{color}
\definecolor{mygreen}{rgb}{0,0.5,0}
\definecolor{mygrey}{rgb}{0.5,0.5,0.5}
\definecolor{myred}{rgb}{0.75,0,0}
\definecolor{myblue}{rgb}{0,0,0.75}
\definecolor{mymagenta}{cmyk}{0,1,0,0.12}
\definecolor{mycyan}{cmyk}{1,0,0,0.12}
\definecolor{myorange}{rgb}{1.,0.5,0}
\definecolor{myviolet}{rgb}{0.6,0.15,0.6}
\definecolor{mybrown}{cmyk}{0,0.50,1,0.41}

\usepackage{ulem}

\newcommand{\subdk}{_{\mathrm{dk}}}
\newcommand{\Dtwo}{\ensuremath{\mathrm{D}_2}{}}

\newcommand{\subcoll}{_{\mathrm{coll}}}
\newcommand{\subwd}{_{\mathrm{wd}}}
\newcommand{\subse}{_{\mathrm{se}}}
\newcommand{\subsd}{_{\mathrm{sd}}}
\newcommand{\subbg}{_{\mathrm{bg}}}
\newcommand{\subasn}{_{\mathrm{ASN}}}

\newcommand{\OD}{\mathcal{D}}
\newcommand{\supLWVC}{^{\mathrm{LWVC}}}
\newcommand{\subLWVC}{_{\mathrm{LWVC}}}
\newcommand{\Ipump}{I_\mathrm{pump}}
\newcommand{\Ppump}{P_\mathrm{pump}}
\newcommand{\omegapump}{\omega_\mathrm{pump}}

\usepackage{siunitx}
\DeclareSIUnit\torr{Torr}
\DeclareSIUnit\amagat{amg}

\makeatletter
\def\@email#1#2{%
 \endgroup
 \patchcmd{\titleblock@produce}
  {\frontmatter@RRAPformat}
  {\frontmatter@RRAPformat{\produce@RRAP{*#1\href{mailto:#2}{#2}}}\frontmatter@RRAPformat}
  {}{}
}%
\makeatother

\newcommand{\ICFO}{ICFO - Institut de Ci\`encies Fot\`oniques, The Barcelona Institute of Science and Technology, 08860 Castelldefels (Barcelona), Spain}
\newcommand{\ICREA}{ICREA - Instituci\'{o} Catalana de Recerca i Estudis Avan{\c{c}}ats, 08010 Barcelona, Spain}
\newcommand{\POLIMI}{Dipartimento di Fisica — Politecnico di Milano, Piazza Leonardo da Vinci 32, 20133 Milano, Italy}
\newcommand{\IFNCNR}{Istituto di Fotonica e Nanotecnologie (IFN) — Consiglio Nazionale delle Ricerche (CNR), Piazza Leonardo da Vinci 32, 20133 Milano, Italy}
\newcommand{\BARI}{Dipartimento Interateneo di Fisica, Universit\'{a}
degli Studi di Bari Aldo Moro, 70126 Bari, Italy}

\begin{document}

\preprint{AIP/123-QED}

\title{
Laser-written micro-channel atomic magnetometer}

\author{Andrea Zanoni}
\affiliation{\POLIMI
}
 \affiliation{\IFNCNR}

\author{Kostas Mouloudakis}%
\affiliation{\ICFO}%

\author{Michael C. D. Tayler}%
\affiliation{\ICFO}%

\author{Giacomo Corrielli}%
\affiliation{\IFNCNR}%

\author{Roberto Osellame}%
\affiliation{\IFNCNR}%

\author{Morgan W. Mitchell}
\affiliation{\ICFO}
\affiliation{\ICREA}

\author{Vito Giovanni Lucivero}
  \affiliation{\ICFO}
  \affiliation{\BARI}

\date{\today}

\begin{abstract}
We demonstrate a sensitive optically-pumped magnetometer using rubidium vapor and \SI{0.75}{\amagat} of nitrogen buffer gas in a sub-mm-width sensing channel excavated by femtosecond laser writing followed by chemical etching. The channel is buried less than 1 mm below the surface of its fused silica host material, which also includes reservoir chambers and micro-strainer connections, to preserve a clean optical environment. Using a zero-field-resonance magnetometry strategy and a sensing volume of 2.25 mm$^3$, we demonstrate a sensitivity of $\approx \SI{1}{\pico\tesla\per\sqrt\hertz}$ at \SI{10}{\hertz}. The device can be integrated with photonic structures and microfluidic channels with 3D versatility. Its sensitivity, bandwidth and stand-off distance will enable detection of localized fields from magnetic nanoparticles and \SI{}{\micro\liter} NMR samples.

\end{abstract}

\maketitle

%


Microfabricated atomic vapor cells  are used in miniaturized atomic devices including frequency references \cite{Hummon2018}, clocks \cite{Knappe2004}, gyroscopes \cite{Cipolletti2021}, Rydberg-atom electrometers \cite{Chen2022} and optically pumped magnetometers (OPMs) 
\cite{Griffith2010, JimenezBook2017}.  In most such devices, the miniaturized cells have \SI{}{\milli\meter} or larger internal vapor dimensions \cite{Jiang2023}. As a rule, larger cell volumes enable better sensitivity but also impose a larger standoff distance from a source to the atoms measuring it \cite{JimenezBook2017}. Measurement of highly localized magnetic fields produced by sources, e.g., micro- or nano-scale electronics, microfluidic nuclear magnetic resonance (NMR) ensembles \cite{Ledbetter2008PNAS,kennedy2014PhDthesis,Eills2022review}, micro-scale biomagnetism \cite{FaivreCR2008, MonteilNM2019} or magnetic micro-and nano-particles \cite{Jofre2023,Xu2006magneticparticles}, could benefit from a monolithic microfluidic platform with sub-\SI{}{\milli\meter}-internal-dimension atomic and fluidic channels. This ``lab-on-chip''  approach to magnetic sensing has been studied with NV-centers \cite{Hoese2021,Allert2022}, and demonstrated tens-of-\SI{}{\micro\tesla\per\sqrt\hertz} sensitivities with \SI{}{\nm} and \SI{}{\um}-scale sample sizes.  With atomic vapors, the microfluidic approach has the potential to reach sub-\SI{}{\pico\tesla\per\sqrt\hertz} sensitivities with \SI{}{\mm}-scale samples \cite{Mitchell2020}.

One proven technique for making both micro-fluidic devices \cite{Bellini2010,Schaap2012, Memeo2021} and atomic vapor cells with arbitrary internal geometries \cite{Lucivero2022} is FLICE (Femtosecond Laser Irradiation
followed by Chemical Etching)\cite{Marcinkevicius2001,Osellame2011}. This technique exploits the nonlinear interaction between glass and focused ultrafast laser pulses to locally increase the material's susceptibility to wet etching processes. FLICE can generate three-dimensional empty channels with arbitrary geometry and micrometric resolution, buried in optical materials such as fused silica. FLICE is moreover compatible with laser-writing of optical waveguides and other optical elements. The combination of these techniques creates a route to integrated devices that simultaneously control optical, fluidic, and atomic elements in a single miniaturized package\cite{Corr21,Pelucchi2022}. 

Prior work has used FLICE to produce a miniaturized vapor cell with a $\SI{1}{\milli\meter}\times\SI{1}{\milli\meter}$ cross section and $\SI{9.5}{\milli\meter}$ vapor length, with a buffer gas density of $\approx \SI{5e-3}{\amagat}$\cite{Lucivero2022}. Due to the low pressure, this cell showed Doppler-free saturated-absorption resonances, of interest for laser frequency stabilization, but also low spin coherence time due to rapid diffusion to the walls and weak optical pumping efficiency, due to radiation trapping\cite{Lucivero2022}. Here we demonstrate a FLICE-made miniaturized vapor cell with a $\SI{500}{\micro\meter}\times \SI{500}{\micro\meter}$  cross-section and \SI{0.75}{\amagat} of N\textsubscript{2} buffer gas. This higher pressure allows us to study the potential of FLICE-made cells for sensitive magnetometry. 

\par

\begin{figure}[t]
\centering
\includegraphics[width=\columnwidth]{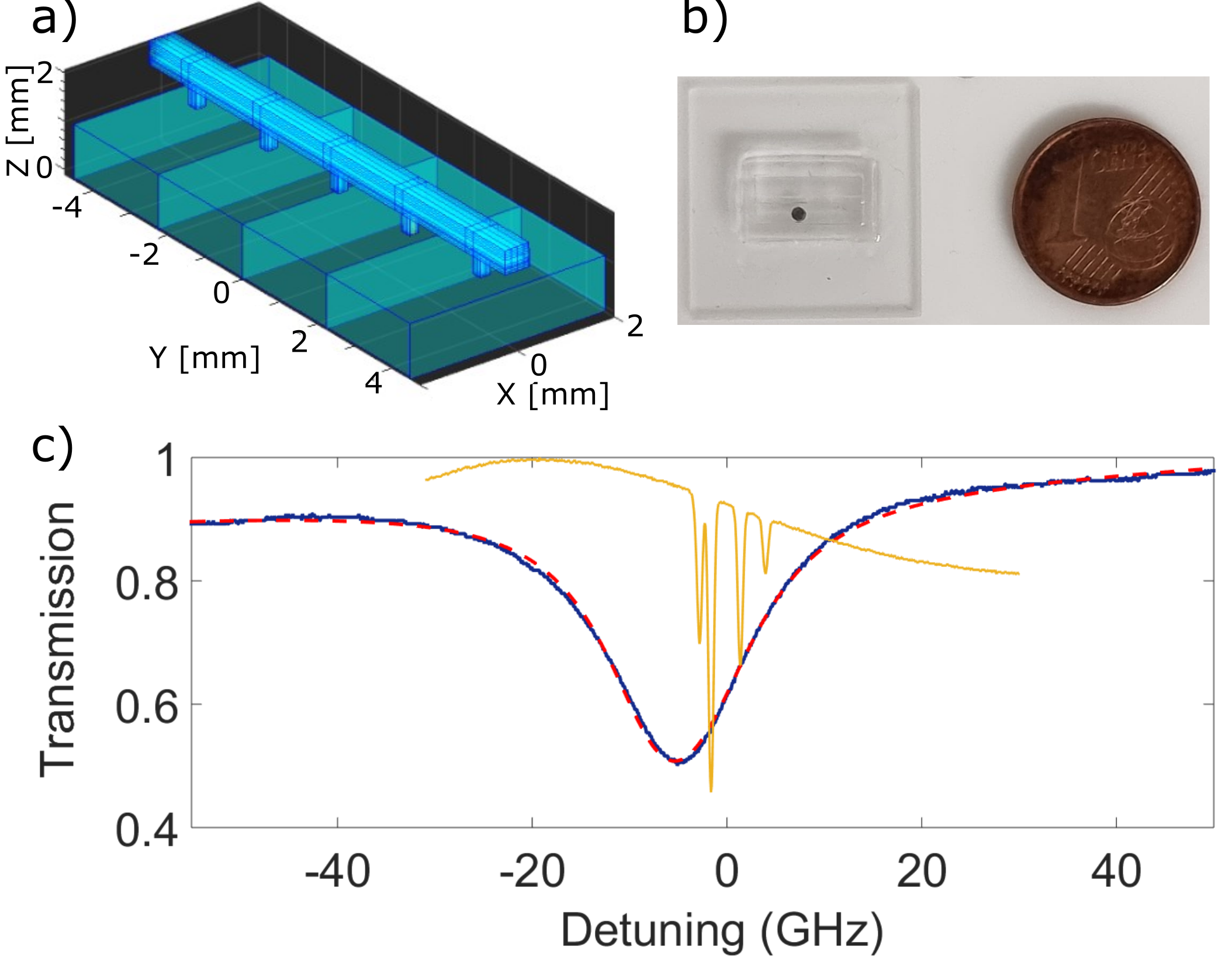}
\caption{ a) Design of the LWVC with a bottom reservoir and a top 9 \si{\milli \meter}-long sensing micro-channel. b) LWVC after fabrication by FLICE, filling with a Rb solid state dispenser and sealing by UV-curing adhesive. c) Normalized transmission (blue) through the LWVC and fit (dashed red) to an absorptive Lorentzian plus a linear dependence on the DBR laser frequency. Simultaneous transmission through a Rb reference vapor cell (yellow) evacuated to 10$^{-8}$ Torr.}
\label{fig:LWVCDesign&SpectrumMod}
\end{figure}
The FLICE process, including filling and bonding steps for the fabrication of laser-written vapor cells (LWVCs), is described in detail in our prior work \cite{Lucivero2022}. In this experiment we use a commercial femtosecond laser (CARBIDE - Light Conversion) to write the desired geometry into a $\SI{20}{\mm}\times \SI{20}{\mm}$ fused silica slab with a thickness of $\SI{3}{\mm}$. As illustrated in Fig. \ref{fig:LWVCDesign&SpectrumMod}(a) the laser-written geometry consists of a sensing micro-channel of length $l\subLWVC=\SI{9}{\mm}$ and side $d\subLWVC=\SI{500}{\um}$ with a square cross section. The distance between the void micro-channel and the silica top surface, which determines the stand-off distance from a potential sample on the top of the on-chip OPM, is \SI{750}{\um} and could be reduced to few tens of microns.  A 4 $\SI{}{\mm}\times$ 8 $\SI{}{\mm}\times$ 1 $\SI{}{\mm}$ bottom reservoir, open at one end, is connected to the top physics channel by five \SI{250}{\um}-wide ``micro strainer'' conduits. This configuration has been designed to give sub-mm confinement of the atomic ensemble over two dimensions with the potential of a sub-mm stand-off distance from a sample, while maintaining high optical depth due to the \SI{9}{\milli\meter} interaction length. The etching process took about \SI{10}{\hour} in hydrofluoric acid (HF) at \SI{20}{\percent} concentration, at \SI{35}{\celsius}, with the acid entering from the open reservoir and reaching the sensing channel to remove the modified material after irradiation and to obtain the final hollow micro structure. Filling and bonding steps have been performed inside a N\textsubscript{2}-filled glove-box at atmospheric pressure. We first placed a dispenser of non-evaporable getter (NEG) material (SAES Getters RB/AMAX/PILL/1-0.6) in the cell reservoir. Then, we used UV-curing epoxy (Norland Products NOA61) to bond the cell to a square fused silica plate of dimensions  $\SI{20}{\mm}\times \SI{20}{\mm}\times \SI{1}{\mm}$. \footnote{We performed leak tests with different epoxy-sealed cells, observing no leaking up to \SI{150}{\celsius}.} A picture of the final cell filled with the Rb dispenser is shown Fig. \ref{fig:LWVCDesign&SpectrumMod}(b). The activation was performed for 20 \si{\second} with a 5 \si{\watt} beam of cw light at 1064 \si{\nano \meter}, reaching the dispenser with a 200 \si{\micro \meter} waist from the top of the cell.  A distributed Bragg reflector (DBR) probe laser was scanned around the center of the  $\Dtwo{}$ line of $^{85}$Rb and detected after transmission through the physics channel to simultaneously monitor the activation by absorption spectroscopy. In Fig. \ref{fig:LWVCDesign&SpectrumMod}(c) we show the absorption spectrum acquired with the cell stabilized at \SI{90}{\celsius} for \SI{2}{\hour} after activation. The absorption spectrum is fitted to a Lorentzian profile to obtain a full-width half maximum (FWHM) pressure-broadened linewidth of 13.5 \si{\giga \hertz}, corresponding to $\eta=0.75$ amg of N$_2$ \cite{Romalis1997}. This residual pressure simultaneously matches two requirements for efficient optical pumping \cite{SeltzerThesis} and high sensitivity optical magnetometry: the N$_2$ works as a quenching gas to avoid radiation trapping  \cite{Rosenberry2007} and as a buffer gas to reduce the rate of depolarizing collisions of the Rb atoms with the cell walls \cite{Lucivero2022,Scholtes2014}. A fine pressure tuning with resolution as low as \SI{1}{\torr} has also been recently demonstrated in a micro-fabricated vapor cell \cite{Dyer2023} using cesium (Cs) dispensers, making the described filling and activation process suitable for chip-scale atomic sensors with different target pressures.\par

\begin{figure}[t]
\centering
\includegraphics[width=\columnwidth]{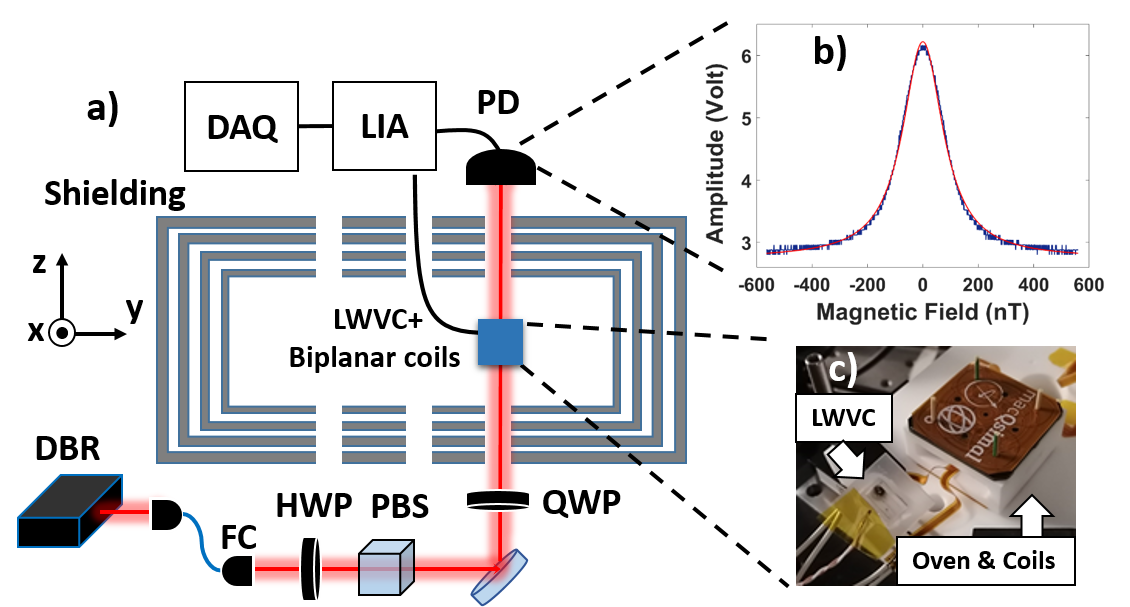}
\caption{a) Experimental setup for optical magnetometry. DBR -  distributed Bragg reflector
laser; FC - Fiber collimator; HWP- half wavepate; PBS - polarizing beam splitter; QWP - quarter waveplate; LWVC - laser-written vapor cell; PD - photodetector; LIA - lock-in amplifier; DAQ - data acquisition system. b) Zero-field resonance (ZFR) signal measured in transmission when a magnetic field is swept around zero in the x-direction, transvese to the laser beam propagation direction z. c) The system comprising the LWVC, a ceramic oven, a teflon insulator, and the biplanar coils. The module is placed within 4 layers of mu-metal magnetic shielding.}
\label{fig:ExpSetup}
\end{figure}

We use the LWVC to implement a zero-field resonance (ZFR) OPM\cite{Dupontroc1969,Shah2007,KrzyzewskiJAP2019}. The experimental setup is shown in Fig. \ref{fig:ExpSetup}(a). 
The light from a $795$ \si{\nano\meter} DBR diode laser, resonant with the $\rm{D}_1$ line of \textsuperscript{87}Rb, is fiber-coupled in a single-mode fiber, whose output is aligned to pass through the atomic sensing channel of the LWVC in the $z$-direction. A half waveplate (HWP) and a polarizing beam splitter (PBS) are used to control the laser power and a zero-order quarter waveplate (QWP) to convert linear into circular polarization before atomic interaction. A spherical lens (not shown) with \SI{100} {\mm} focal length is used to shape the laser beam to have a waist of \SI{250}{\um} at the center of the LWVC. The cell was placed in a ceramic oven, heated through ac current at \SI{40}{\kHz} through flex-PCB resistive heater traces attached to the ceramic box. The setup is thermally insulated using a Teflon enclosure. Uniform magnetic fields along the $x,y$ and $z$ directions were generated by a set of four flexible-PCB coils implemented with a biplanar design \cite{Tayler2022}. The coils were attached on top and bottom faces of the insulation box with an inter-plane distance of 16 \si{\milli\meter}. The wire paths were designed using the \texttt{bfieldtools} software \cite{bfieldtools2} and  optimized for high field homogeneity and low stray field. Both the Teflon and the ceramic parts have 3 \si{\milli \meter} wide windows allowing optical access to the cell. The complete LWVC and oven package, shown in Fig. \ref{fig:ExpSetup}(c), is placed inside four cylindrical layers of mu-metal magnetic shielding. 
Transmitted light is focused onto an amplified Si photodetector (PD). 

When a magnetic field is swept around zero in the x-direction, a magnetic ZFR, as the one shown in Fig. \ref{fig:ExpSetup}(b), is obtained. To obtain a signal linear around zero applied field, we sinusoidally modulate the field in the x-direction and demodulate the PD signal using a lock-in amplifier (LIA). The LIA (SRS SR830) quadrature output is digitized by a DAQ with $5$MS/s sample rate for an acquisition time of $t=0.5$ sec. Two coils are driven by a low-noise current source (Twinleaf CSB-10) to generate dc fields in the $y,z$ directions to cancel the ambient residual field in the magnetic shield, which is a few tens of \si{\nano\tesla}, and to maximize the ZFR sharpness. 

\begin{figure}[t!]
\centering
\includegraphics[width=\columnwidth]{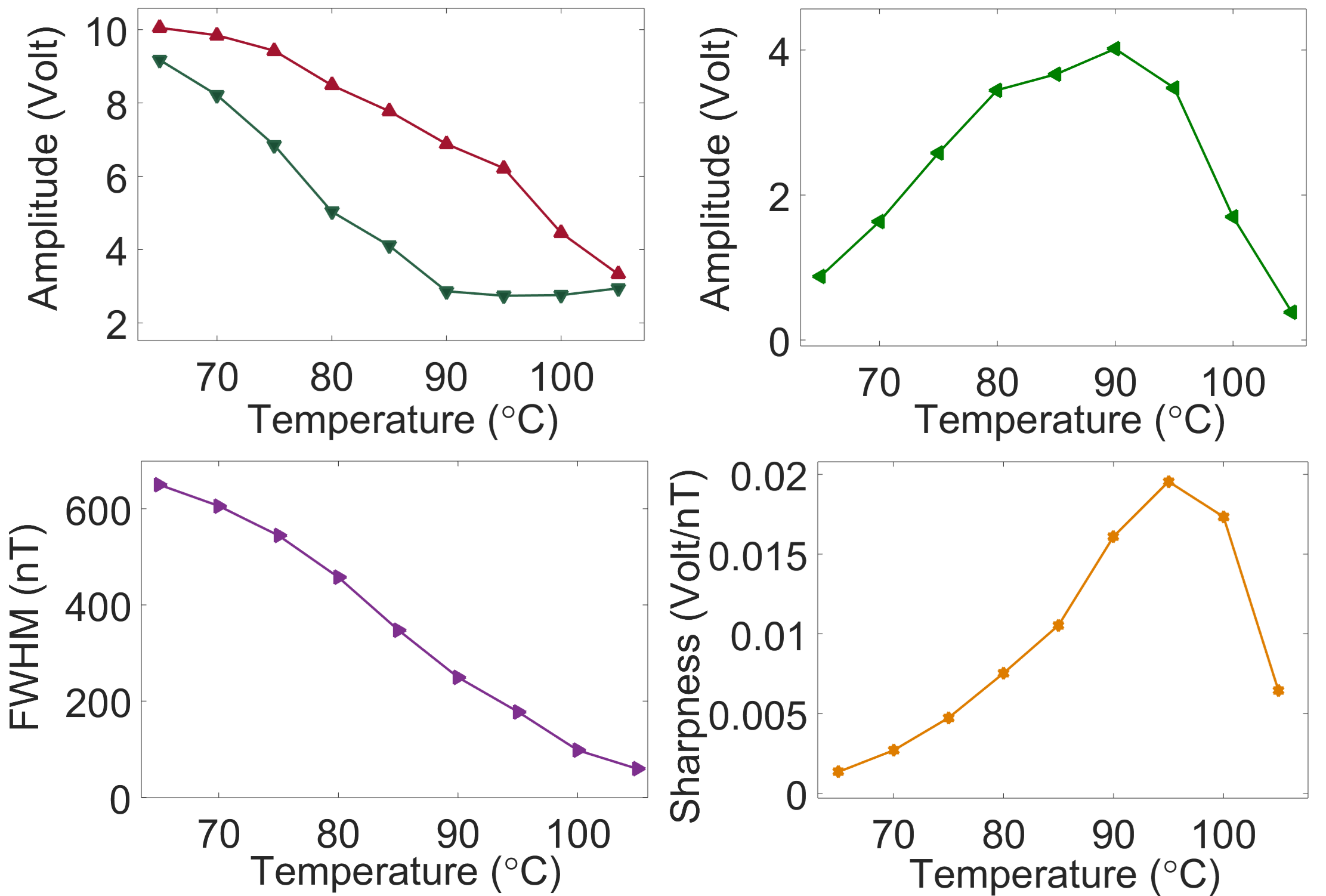}
\caption{ZFR parameters versus LWVC temperature, acquired with pump power of \SI{360}{\micro\watt}. C.f. Fig.~\ref{fig:ExpSetup} and Eq.~(\ref{eq:VofBLorentzianFitFunction}). (upper left) photodetector voltage minimum $a$ (green) and maximum $b$ (red)  b). (upper right)  amplitude $b-a$. (lower left) FWHM linewidth $\Delta B$. (lower right)  sharpness $s = (b-a)/\Delta B$, showing an optimum at \SI{96}{\celsius}.}
\label{fig:ZFRAmplitudes}
\end{figure}

As described in prior works on single-beam zero-field magnetometry \cite{Dupontroc1969,Shah2007,KrzyzewskiJAP2019}, we apply a DC magnetic field component $B_x=\gamma^{-1}\omega_L\hat{x}$, where $\gamma$ is the gyromagnetic ratio and $\omega_L$ is the Larmor frequency, and a parallel oscillating magnetic field $B_m=\gamma^{-1}\omega_1\cos(\omega_mt)\hat{x}$, with amplitude $\gamma^{-1}\omega_1$ and frequency $\omega_m$, in the x-direction, transverse to the laser beam propagation. As described in Appendix \ref{AppendixA}, the magnetic-field-dependent changes of the degree of spin polarization on-axis $P_z$ are mapped onto the transmitted intensity, which is converted by the photodetector into an output voltage $V(B)$. This signal, as illustrated in Fig.~\ref{fig:ExpSetup}(b), is well fit by 
the Lorentzian
\begin{equation}
    V(B)=a+(b-a)\frac{\Delta B^2/4}{(B_x-B_0)^2+\Delta B^2/4},
    \label{eq:VofBLorentzianFitFunction}
\end{equation}  
where $a$ is the minimum, $b$ is the maximum, $(b-a)$ is the amplitude, $B_0$ is the line center that may differ from zero in case of residual uncompensated field, and $\Delta B$ is the FWHM linewidth. We use the observed amplitude and linewidth to optimize the sharpness, defined as $s = (b-a)/\Delta B$, by choice of cell temperature. As shown in Fig. \ref{fig:ZFRAmplitudes}, by increasing the temperature and thus the number density, the amplitude increases due to the higher number of atoms contributing to the ZFR signal up to a maximum value. Above this optimum, we observe a net decrease in the maximum voltage as well as in the ZFR amplitude while the FWHM linewidth decreases with number density,  such that an optimum of the sharpness is obtained at \SI{96}{\celsius}. In the low-optical-depth regime, $\Delta B$ can be related to the total magnetic relaxation rate $\Gamma = \gamma \Delta B = R_{\rm{op}}+\Gamma\subdk$ \cite{CastagnaPRA2011}, where $R_{\rm{op}}$ is the optical pumping rate and $\Gamma\subdk$ is the relaxation rate ``in the dark''. Here, however, the optical depth at \SI{96}{\celsius} is $\OD_0=0.9$, calculated by Eq. \ref{Eq:OpticalDepth} in Appendix \ref{AppendixA}, the atomic polarization inhomogeneous across the cell and the relaxation-linewidth relation is expected to be more complex.

\begin{figure}[t!]
\centering
\includegraphics[width=\columnwidth]{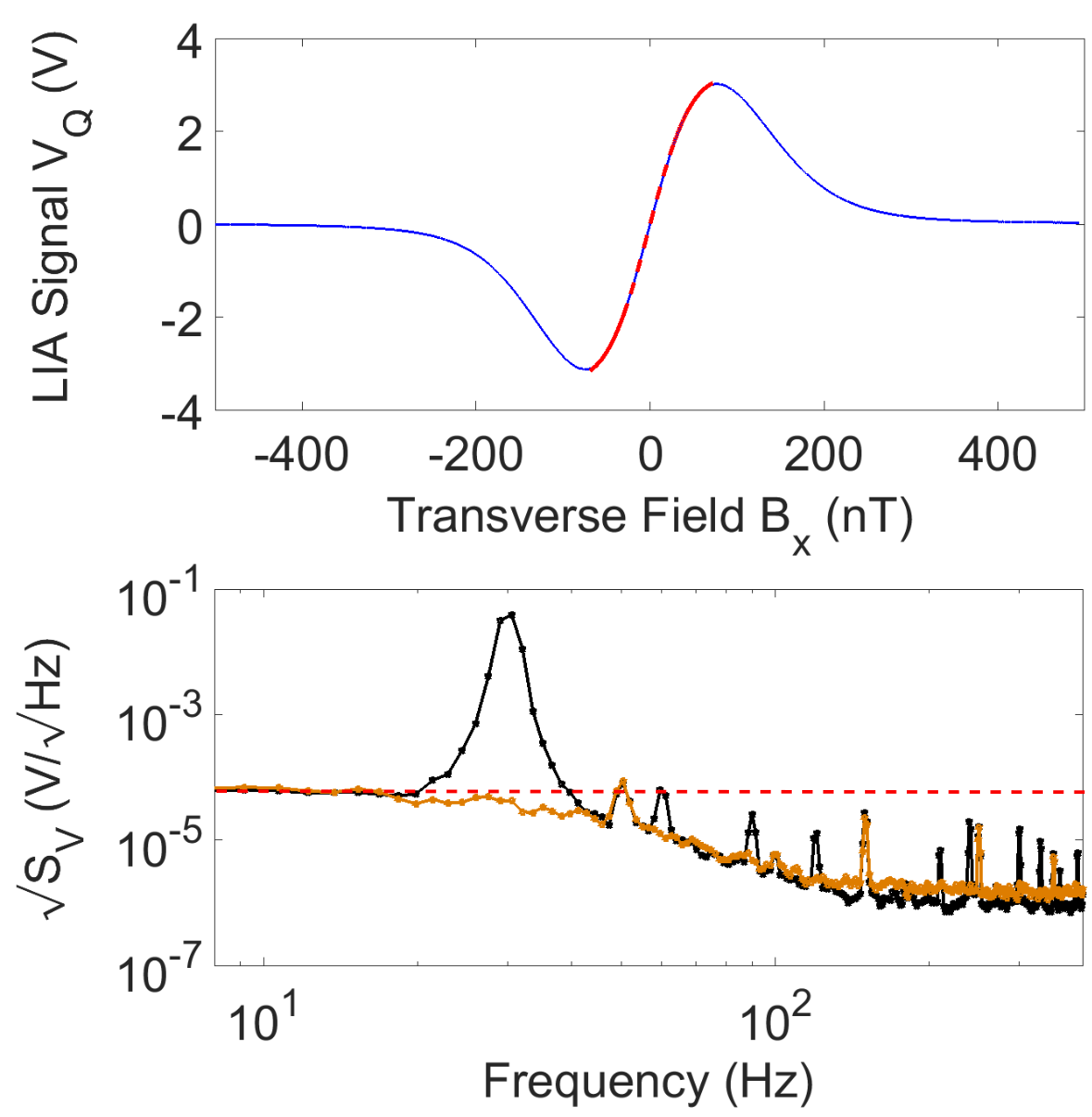}
\caption{LIA quadrature output signal and noise. (Top) Experimental quadrature signal $V_Q(B)$ (blue) versus transverse magnetic field $B_x$ and fit (red) to Eq. \ref{eq:DispersiveLorentzian}. (Bottom) Amplitude spectral density (ASD) of the LIA output voltage at the zero crossing condition with (black) and without (orange) a \SI{30}{\hertz} reference signal. The LIA noise $S_V$, converted to equivalent magnetic noise using the slope of the LIA signal, implies a magnetic sensitivity of \SI{1}{\pico\tesla\per\sqrt\hertz}, shown by the red dashed line, for frequencies below the LIA filter roll-off (see text).
} 
\label{fig:Quad&Sens}
\end{figure}

As illustrated in Fig. \ref{fig:ExpSetup}(a), we use a LIA to demodulate the photodetector output at the same modulation frequency $\omega_m$ of the applied oscillating field $B_m$. In Fig. \ref{fig:Quad&Sens} (top) we show the experimental LIA quadrature output voltage for a quasi-static scan, under optimal ZFR sharpness conditions at \SI{96}{\celsius}, and a fit to a dispersive Lorentzian profile
\begin{equation}
V_Q(B)=\frac{u}{2}\frac{(B-B_0)\Delta B}{(B-B_0)^2+\Delta B^2/4}
\label{eq:DispersiveLorentzian}
\end{equation}
with amplitude $u$. From the fit shown in Fig. \ref{fig:Quad&Sens} (top) we obtain an amplitude of $u=$ 6.4 $\SI{}{\volt}$ and a FWHM linewidth of $\Delta B=182$ \si{\nano \tesla}.  In Fig. \ref{fig:Quad&Sens} (bottom) we show the amplitude spectral density (ASD) obtained in $\SI{}{\volt\per\sqrt\hertz}$ with a FFT of the LIA signal in the time domain fed into a DAQ at the zero-crossing condition, as well as the same spectrum with a reference magnetic signal oscillating at 30 Hz with an rms amplitude of 61 $\SI{}{\pico\tesla}$. As in \cite{Gerginov2017,JimenezMartinez2012,Troullinou2023}, at given detection frequency we can obtain 
the equivalent magnetic noise (power spectral density), as:
\begin{equation}
\label{eq:MagSens}
S_B = {\Big(\frac{dV_Q}{dB}\Big)^{-2}S_V}={\Big(\frac{\Delta B}{2u}\Big)^{2}S_V}
\end{equation}
where $dV_Q/dB$ is the slope of the LIA quadrature signal and $S_V$ is the power spectral noise density in $\si{\volt}^2/\si{\hertz}$. We use the maximum slope, around resonance $B=B_0\approx 0$, which is equal to $dV_Q/dB=2u/\Delta B$.
At \SI{10}{\hertz} detection frequency we found a noise level about $\SI{1}{\pico\tesla\per\sqrt\hertz}$, calculated by using Eq. \ref{eq:MagSens} with $S_{V}$ from Fig.~\ref{fig:Quad&Sens} (bottom) and slope from the fit in Fig.~\ref{fig:Quad&Sens} (top).  
The roll-off of the $S_V$ is due to a LIA 24 dB/oct low-pass filter with 1 \SI{}{\milli\sec} time constant. We then expect a 3 dB measurement bandwidth of the OPM limited to a cutoff frequency of $\approx \SI{160}{\hertz}$, given that the calculated relaxation rates (see Appendix \ref{AppendixA}), and thus the intrinsic bandwidth of the ZFR process, are much faster.

\par
We have described a sensitive single-beam zero-field OPM using a LWVC with a \SI{9}{\mm}-long Rb-vapor-filled channel of $\SI{500}{\micro\meter} \times \SI{500}{\micro\meter}$ cross section, buried less than \SI{1}{\mm} below the surface of the device. With a zero-field-resonance technique, we demonstrate a sensitivity of $\SI{1}{\pico\tesla\per\sqrt\hertz}$ at \SI{10}{\hertz} that, in combination with low-frequency bandwidth and a potential sub-mm standoff distance, makes the device suitable for lab-on-chip magnetometry applications \cite{Allert2022}. These include NMR with OPMs \cite{Ledbetter2013} for quantitative chemicals analysis and imaging in the zero-to-ultralow-field (ZULF) regime \cite{Bajaj2010,Jimenez-MartinezNC2014,Ledbetter2013,Theis2011, ChuchkovaJPCL2023, Mouloudakis2023}, plus detection of other systems such as magnetic nanoparticles \cite{Jofre2023} and magnetotactic bacteria (MTB) \cite{Khalil2013}. Possible damage of bio-samples due to operating temperatures of Rb could be reduced using Cs vapors or metastable He as well as adopting efficient on-chip thermal isolation methods \cite{Ceccarelli2019}. The relaxation rate due to collisions with the walls could be further reduced by a higher buffer gas pressure \cite{Ciriolo2020} or by anti-relaxation coatings \cite{Vasilakis2015}. The FLICE approach to vapor cell production also enables integration of miniature vapor cells with photonic structures, such as waveguides \cite{Hummon2018}.

\begin{acknowledgments}
Work supported by: Spanish Ministry of Science MCIN project SAPONARIA (PID2021-123813NB-I00) and ``Severo Ochoa'' Center of Excellence CEX2019-000910-S; Departament de Recerca i Universitats de la Generalitat de Catalunya grant No. 2021 SGR 01453;  Fundaci\'{o} Privada Cellex; Fundaci\'{o} Mir-Puig; 
GC, RO, and VGL acknowledge financial support from European Union NextGenerationEU (PNRR MUR project PE0000023 -- NQSTI Spoke 7).  MT acknowledges projects MARICHAS (PID2021-126059OA-I00) and RYC2022-035450-I financed by MCIN/ AEI /10.13039/501100011033. KM acknowledges support from Grant FJC2021-047840-I funded by MCIN/AEI/ 10.13039/501100011033 and by the European Union project ``NextGenerationEU/PRTR.'' 
Funded by the European Union (ERC, Field-Seer, 101097313). Views and opinions expressed are however those of the author(s) only and do not necessarily reflect those of the European Union or the European Research Council Executive Agency. Neither the European Union nor the granting authority can be held responsible for them.

\end{acknowledgments}

\section*{Data Availability Statement}
The dataset that support the findings of this study are openly available in the repository \url{https://zenodo.org/records/10974079}.

\appendix

\section{optical depth and collisional rates}
\label{AppendixA}
The intensity $I$ of the pump beam, as it propagates through the laser-written atomic channel along the z-direction, depends on the on-axis degree of self-induced spin polarization $P_z \in [-1,1]$ via the Beer-Lambert law:
\begin{equation}
\frac{dI(z)}{dz} = -n \sigma(\nu) I(z)[(1-  P_z (z)]
\label{eq:BeerLambert}
\end{equation}
where $n$ is the Rb number density, $\sigma(\nu)=\pi r_ecf_{osc}^{D1}\mathcal{L}(\nu-\nu_0)$ is the absorption cross-section, $r_e$ is the classical electron radius, $c$ is the speed of light, $f_{osc}^{D1}\approx 0.3423$, $\mathcal{L}(\nu-\nu_0)$ is a normalized Lorentzian absorption profile with pressure-broadened FWHM linewidth $\Gamma_L$. For simplicity, we consider a homogeneous atomic polarization by neglecting the spatial distribution of $P_z (z)$ across the laser-written channel and the consequent integral over the cell length \cite{Ito2016}. In this qualitative approach, the transmitted intensity of the pump beam, tuned near the central line of $^{85}$Rb to maximize the ZFR amplitude, that reaches the photodetector is:
\begin{equation}
\label{eq:TransmittedIntensity}
I=I_0\mathrm{exp}[-n\sigma_0l(1-P_z)]=I_0\mathrm{exp}[-\OD_0(1-P_z)]
\end{equation}
where $I_0$ is the intensity at the input of the cell and the on-resonance optical depth is given by: 
\begin{equation}
\label{Eq:OpticalDepth}
\OD_0 = \frac{\pi n r_e c f_{osc}^{D1} l}{\Gamma_L/2} 
\end{equation}
where $l$ is the atomic interaction length. With $l=l\subLWVC$, $T=\SI{96}{\celsius}$ and $\Gamma_L=2\pi\times13.5$ GHz, we calculate $\OD_0=0.9$.

\begin{figure}[t]
\centering
\includegraphics[width=\columnwidth]{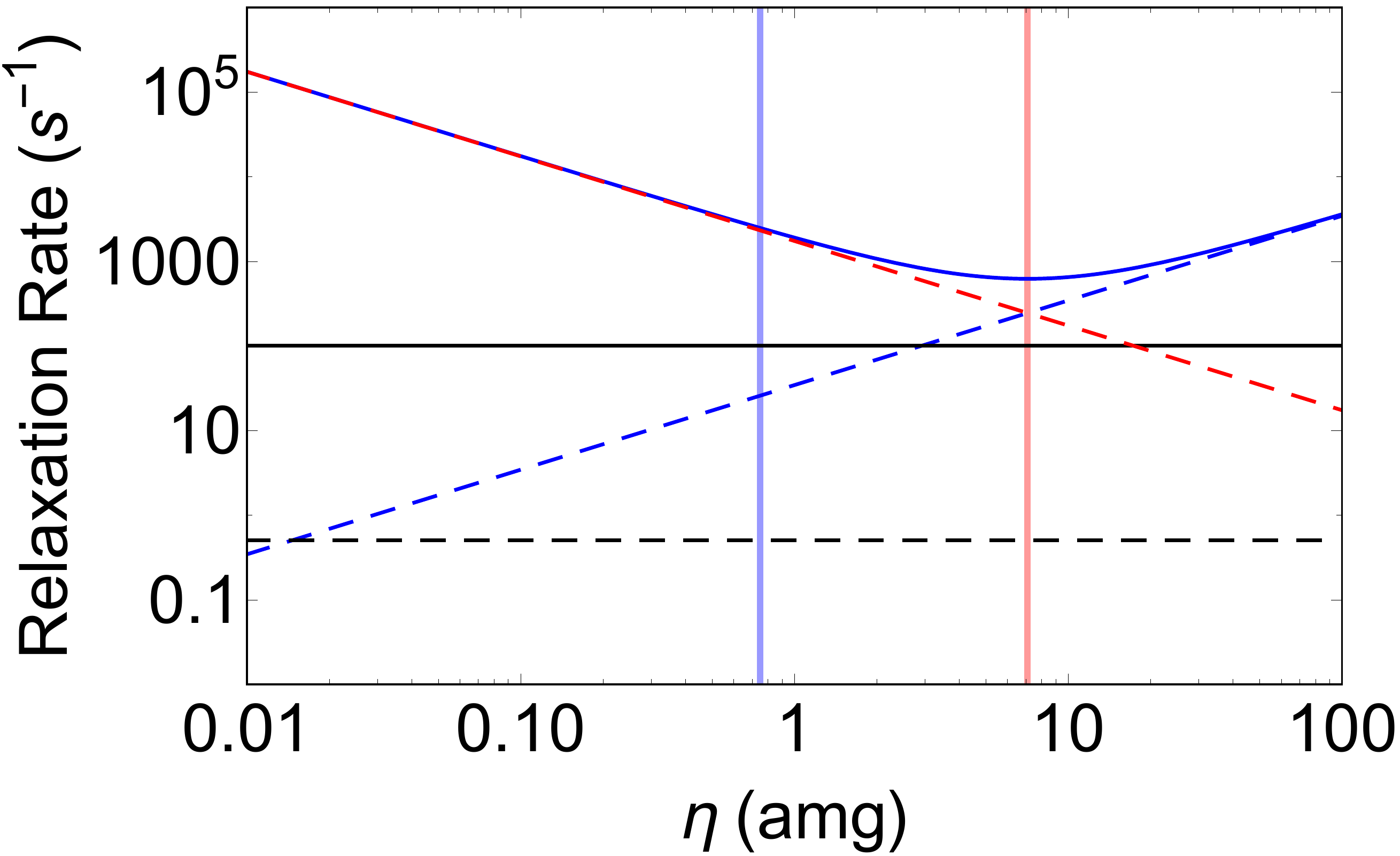}
\caption{Collisional relaxation rates as a function of N$_2$ buffer gas density at $T=\SI{96}{\celsius}$. $\Gamma\subwd\supLWVC$ (dashed red), $\Gamma\subbg$ (dashed blue), $\Gamma\subsd$ (dashed black), $\Gamma\subse$ (continuous black), total $\Gamma\subdk$ (continuous blue). Experimental (optimal) nitrogen density is indicated with a vertical blue (red) line at 0.75 amg (7.1 amg).}
\label{fig:RelRatesCalc}
\end{figure}

In our experimental condition, with enough buffer gas to prevent radiation trapping \cite{Rosenberry2007}, the relaxation rate in the dark is given by:
\begin{equation}
\Gamma\subdk=\Gamma\subwd+\Gamma\subcoll,
\label{eq:RelRate}
\end{equation}
where $\Gamma\subwd$ and $\Gamma\subcoll$ are the depolarizing rates due to atomic diffusion to the walls and to binary collisions. The relaxation rate $\Gamma\subwd$ for the fundamental diffusion mode of a rectangular cell is \cite{Kitching2018}:
\begin{equation}
\Gamma\subwd\supLWVC=\Big[\Big(\frac{\pi}{l_x}\Big)^2+\Big(\frac{\pi}{l_y}\Big)^2 +\Big(\frac{\pi}{l_z}\Big)^2\Big]\frac{D_0}{\eta}\sqrt{\frac{T(\SI{}{\kelvin})}{273.15\SI{}{\kelvin}}},
\label{eq:WallsRateLWVC}
\end{equation}
where $l_x=l_y=d\subLWVC$ and $l_z=l\subLWVC$ are the cavity side lengths, $D_0=\SI{0.12}{\centi\meter\squared\per\s}$ is the diffusion constant  in \SI{1}{\amagat} of N\textsubscript{2} buffer gas  at \SI{273}{\kelvin} \footnote{Obtained from the value $D_0=\SI{0.16}{\centi\meter\squared\per\second}$ measured in \cite{IshikawaPRA2000} at $T_0=\SI{60}{\celsius}$  using the scaling $D \propto (T(\SI{}{\kelvin})/T_0(\SI{}{\kelvin}))^{3/2}$ \cite{Lucivero2017}.} and $\eta$ is the the nitrogen number density in \SI{}{\amagat}. 
The collisional relaxation rate is $\Gamma\subcoll=\Gamma\subbg+\Gamma\subse+\Gamma\subsd$, including Rb-buffer gas ($\Gamma\subbg$), Rb-Rb spin-exchange  ($\Gamma\subse$) and Rb-Rb spin-destruction ($\Gamma\subsd$) collisional rates, respectively:
\newcommand{\Rb}{\mathrm{Rb}}
\newcommand{\Ntwo}{\mathrm{N}_2}
\newcommand{\dash}{\text{-}}
\newcommand{\se}{\mathrm{se}}
\renewcommand{\sd}{\mathrm{sd}}
\begin{eqnarray}
\label{eq:se} \Gamma\subse&=&q_{\se}n\sigma_{\se}\bar{v}_{\Rb\dash\Rb},  \\ \label{eq:sd}
\Gamma\subsd&=&n\sigma_{\sd}\bar{v}_{\Rb\dash\Rb}, \\ \label{eq:buffgas}
\Gamma\subbg&=&n_{N_2}\sigma_{\Rb\dash\Ntwo}\bar{v}_{\Rb\dash\Ntwo},
\end{eqnarray}
where $n$ ($n_{N_2}$) is the Rb (N$_2$) number density, assuming nitrogen as an ideal gas, $\bar{v}_{\Rb\dash\Ntwo}$ ($\bar{v}_{\Rb\dash\Rb}$) is the Rb-N\textsubscript{2} (Rb-Rb) relative thermal velocity, $q_{\se}=5/27$ is the spin-exchange broadening factor \footnote{For simplicity we neglect spin-exchange collisions between the two Rb isotopes and we use the spin-exchange broadening factor for $^{85}$Rb, obtained with nuclear spin $I=5/2$ in \cite{SeltzerThesis}}, $\sigma_{\se}=\SI{1.9e-14}{\centi\meter\squared}$, $\sigma_{\sd}=\SI{1.6e-17}{\centi\meter\squared}$ and $\sigma_{\Rb\dash\Ntwo}=\SI{1.e-22}{\centi\meter\squared}$ are the cross-sections for Rb-Rb spin-exchange and spin-destruction collisions and Rb-N\textsubscript{2} spin destruction collisions, respectively\cite{SeltzerThesis}. In Fig. \ref{fig:RelRatesCalc} we show the nitrogen-density-dependent rates from Eqs. \ref{eq:WallsRateLWVC} and \ref{eq:buffgas}, the nitrogen-density-independent rates given by Eqs. \ref{eq:se} and \ref{eq:sd}, and the total collisional relaxation rate using Eq. \ref{eq:RelRate}, all calculated at $T=\SI{96}{\celsius}$. At our experimental density of $\eta=0.75$ amg, we obtain a dominant contribution due to collisions with the walls $\Gamma\subwd\supLWVC=2\pi\times2326$ \SI{}{\per\second}, a spin-exchange rate $\Gamma\subse=2\pi\times111.2$ \SI{}{\per\second}, Rb-buffer gas and spin-destruction rates $\Gamma\subbg=2\pi\times25.7$ \SI{}{\per\second} and $\Gamma\subsd=2\pi\times0.5$ \SI{}{\per\second}, respectively, and a total rate in the dark of $\Gamma\subdk=2\pi\times2483$ \SI{}{\per\second}. An optimum of $\eta_{opt}=7.1$ amg would correspond to a minimum total rate of $\Gamma\subdk=2\pi\times600$ \SI{}{\per\second}. The on-resonance local rate of optical pumping is:
\begin{equation}
R_{\rm{op}}=\sigma_0\phi_\mathrm{phot}=
\frac{r_e c f_{osc}^{D1}}{\Gamma_L/2} \frac{\Ipump}{\hbar \omegapump}
\label{eq:PumpRate}
\end{equation}
where $\omegapump=2\pi\times3.77\times10^{14}$ \SI{}{\hertz}, $\phi_\mathrm{phot}$ is the photon flux, $\Ipump=\Ppump/(\pi w^2/2)$ is the pump intensity for a Gaussian beam with radius $w$ and power $\Ppump$. Before atomic interaction and beam focusing into the sensing channel, we measured a power of \SI{360}{\micro\watt} for a beam radius of $w\sim\SI{1}{\milli\meter}$. However, given a room-temperature experimental transmission of about $15\%$ through the laser-written channel, in Eq. \ref{eq:PumpRate} we use $\Ppump=\SI{55}{\micro\watt}$ and we obtain $R_{\rm{op}}=2\pi\times1513$ \SI{}{\per\second}. This rate is overestimated due to the assumption of homogeneous polarization and pump intensity across the cell in Eq. \ref{eq:PumpRate}.

\section{projected sensitivity versus sensing volume}
\label{AppendixB}

While technical and optical noise sources often prevent to achieve atomic-noise-limited sensor performance, in the context of miniaturized atomic sensors, as the described LWVC with sub-mm dimensions, we can gain insights on the dependence on sensing volume $V$, by calculating the projected magnetic sensitivity due to atomic shot noise (ASN) \cite{Sheng2013,Shah2007}:
\begin{equation}
\delta B\subasn=\frac{1}{\gamma}\sqrt{\frac{2e\Gamma\subdk}{N_{at}t}}=\frac{1}{\gamma}\sqrt{\frac{2e\Gamma\subdk}{nVt}}
\label{eq:ASN}
\end{equation}
where $N_{at}=nV$ is the total number of atoms and t is the measurement time. By using the relaxation rate in the dark $\Gamma\subdk$, already defined in Eq. \ref{eq:RelRate}, we neglect optically-induced relaxation effects, which can be significantly reduced in other types of sensitive magnetometers, as those based on free-induction-decay (FID) and off-resonance probing \cite{Gerginov2017, Limes2020}. At given operating temperature $T$ and volume, thus fixed geometrical parameters in the rate due to diffusion to the walls $\Gamma\subwd$, one can find the buffer gas density that minimizes both $\Gamma\subdk$ and the projected sensitivity through Eq. \ref{eq:ASN}. This value typically corresponds to the crossing point between $\Gamma\subwd$, which decreases with increased N$_2$ pressure and $\Gamma\subbg$, which in contrast increases with the nitrogen density, as shown in Fig. \ref{fig:RelRatesCalc}. For the sensing volume of the laser-written channel used in this experiment $V\subLWVC=\mathrm{d\subLWVC}^2\times\mathrm{l\subLWVC}=2.25$ mm$^3$, and working temperature of $T=\SI{96}{\celsius}$, optimal for the ZFR sharpness, the optimal buffer gas density corresponds to $\eta_{opt}=$7.1 amg, which gives a projected magnetic sensitivity of 18.3 $\SI{}{\femto\tesla\per\sqrt\hertz}$, calculated by using Eq. \ref{eq:ASN} with $t=0.5$ sec. As illustrated in Fig. \ref{fig:RelRatesCalc}, for our experimental condition of $\eta=\SI{0.75}{\amagat}$ the relaxation rate due to alkali-walls collisions $\Gamma\subwd\supLWVC$ is nonetheless significantly decreased, resulting in a projected ASN-limited sensitivity of 36.6 $\SI{}{\femto\tesla\per\sqrt\hertz}$. In practice, ZFR-OPMs as the described magnetometer, present additional optical and technical noise \cite{Shah2007,KrzyzewskiJAP2019}.





\bibliographystyle{apsrev4-1}
\bibliography{lwvcopm}

\providecommand{\noopsort}[1]{}\providecommand{\singleletter}[1]{#1}%
\begin{thebibliography}{58}%
\makeatletter
\providecommand \@ifxundefined [1]{%
 \@ifx{#1\undefined}
}%
\providecommand \@ifnum [1]{%
 \ifnum #1\expandafter \@firstoftwo
 \else \expandafter \@secondoftwo
 \fi
}%
\providecommand \@ifx [1]{%
 \ifx #1\expandafter \@firstoftwo
 \else \expandafter \@secondoftwo
 \fi
}%
\providecommand \natexlab [1]{#1}%
\providecommand \enquote  [1]{``#1''}%
\providecommand \bibnamefont  [1]{#1}%
\providecommand \bibfnamefont [1]{#1}%
\providecommand \citenamefont [1]{#1}%
\providecommand \href@noop [0]{\@secondoftwo}%
\providecommand \href [0]{\begingroup \@sanitize@url \@href}%
\providecommand \@href[1]{\@@startlink{#1}\@@href}%
\providecommand \@@href[1]{\endgroup#1\@@endlink}%
\providecommand \@sanitize@url [0]{\catcode `\\12\catcode `\$12\catcode
  `\&12\catcode `\#12\catcode `\^12\catcode `\_12\catcode `\%12\relax}%
\providecommand \@@startlink[1]{}%
\providecommand \@@endlink[0]{}%
\providecommand \url  [0]{\begingroup\@sanitize@url \@url }%
\providecommand \@url [1]{\endgroup\@href {#1}{\urlprefix }}%
\providecommand \urlprefix  [0]{URL }%
\providecommand \Eprint [0]{\href }%
\providecommand \doibase [0]{http://dx.doi.org/}%
\providecommand \selectlanguage [0]{\@gobble}%
\providecommand \bibinfo  [0]{\@secondoftwo}%
\providecommand \bibfield  [0]{\@secondoftwo}%
\providecommand \translation [1]{[#1]}%
\providecommand \BibitemOpen [0]{}%
\providecommand \bibitemStop [0]{}%
\providecommand \bibitemNoStop [0]{.\EOS\space}%
\providecommand \EOS [0]{\spacefactor3000\relax}%
\providecommand \BibitemShut  [1]{\csname bibitem#1\endcsname}%
\let\auto@bib@innerbib\@empty
\bibitem [{\citenamefont {Hummon}\ \emph {et~al.}(2018)\citenamefont {Hummon},
  \citenamefont {Kang}, \citenamefont {Bopp}, \citenamefont {Li}, \citenamefont
  {Westly}, \citenamefont {Kim}, \citenamefont {Fredrick}, \citenamefont
  {Diddams}, \citenamefont {Srinivasan}, \citenamefont {Aksyuk},\ and\
  \citenamefont {Kitching}}]{Hummon2018}%
  \BibitemOpen
  \bibfield  {author} {\bibinfo {author} {\bibfnamefont {M.~T.}\ \bibnamefont
  {Hummon}}, \bibinfo {author} {\bibfnamefont {S.}~\bibnamefont {Kang}},
  \bibinfo {author} {\bibfnamefont {D.~G.}\ \bibnamefont {Bopp}}, \bibinfo
  {author} {\bibfnamefont {Q.}~\bibnamefont {Li}}, \bibinfo {author}
  {\bibfnamefont {D.~A.}\ \bibnamefont {Westly}}, \bibinfo {author}
  {\bibfnamefont {S.}~\bibnamefont {Kim}}, \bibinfo {author} {\bibfnamefont
  {C.}~\bibnamefont {Fredrick}}, \bibinfo {author} {\bibfnamefont {S.~A.}\
  \bibnamefont {Diddams}}, \bibinfo {author} {\bibfnamefont {K.}~\bibnamefont
  {Srinivasan}}, \bibinfo {author} {\bibfnamefont {V.}~\bibnamefont {Aksyuk}},
  \ and\ \bibinfo {author} {\bibfnamefont {J.~E.}\ \bibnamefont {Kitching}},\
  }\href {\doibase 10.1364/OPTICA.5.000443} {\bibfield  {journal} {\bibinfo
  {journal} {Optica}\ }\textbf {\bibinfo {volume} {5}},\ \bibinfo {pages} {443}
  (\bibinfo {year} {2018})}\BibitemShut {NoStop}%
\bibitem [{\citenamefont {Knappe}\ \emph {et~al.}(2004)\citenamefont {Knappe},
  \citenamefont {Shah}, \citenamefont {Schwindt}, \citenamefont {Hollberg},
  \citenamefont {Kitching}, \citenamefont {Liew},\ and\ \citenamefont
  {Moreland}}]{Knappe2004}%
  \BibitemOpen
  \bibfield  {author} {\bibinfo {author} {\bibfnamefont {S.}~\bibnamefont
  {Knappe}}, \bibinfo {author} {\bibfnamefont {V.}~\bibnamefont {Shah}},
  \bibinfo {author} {\bibfnamefont {P.~D.~D.}\ \bibnamefont {Schwindt}},
  \bibinfo {author} {\bibfnamefont {L.}~\bibnamefont {Hollberg}}, \bibinfo
  {author} {\bibfnamefont {J.}~\bibnamefont {Kitching}}, \bibinfo {author}
  {\bibfnamefont {L.-A.}\ \bibnamefont {Liew}}, \ and\ \bibinfo {author}
  {\bibfnamefont {J.}~\bibnamefont {Moreland}},\ }\href
  {https://pubs.aip.org/aip/apl/article-abstract/85/9/1460/116970/A-microfabricated-atomic-clock?redirectedFrom=fulltext}
  {\bibfield  {journal} {\bibinfo  {journal} {Applied Physics Letters}\
  }\textbf {\bibinfo {volume} {85}},\ \bibinfo {pages} {1460} (\bibinfo {year}
  {2004})}\BibitemShut {NoStop}%
\bibitem [{\citenamefont {Cipolletti}\ \emph {et~al.}(2021)\citenamefont
  {Cipolletti}, \citenamefont {Riedrich-Moeller}, \citenamefont {Fuchs},
  \citenamefont {Wickenbrock},\ and\ \citenamefont {Budker}}]{Cipolletti2021}%
  \BibitemOpen
  \bibfield  {author} {\bibinfo {author} {\bibfnamefont {R.}~\bibnamefont
  {Cipolletti}}, \bibinfo {author} {\bibfnamefont {J.}~\bibnamefont
  {Riedrich-Moeller}}, \bibinfo {author} {\bibfnamefont {T.}~\bibnamefont
  {Fuchs}}, \bibinfo {author} {\bibfnamefont {A.}~\bibnamefont {Wickenbrock}},
  \ and\ \bibinfo {author} {\bibfnamefont {D.}~\bibnamefont {Budker}},\ }in\
  \href {\doibase 10.1109/SENSORS47087.2021.9639519} {\emph {\bibinfo
  {booktitle} {2021 IEEE Sensors}}}\ (\bibinfo {year} {2021})\ pp.\ \bibinfo
  {pages} {1--4}\BibitemShut {NoStop}%
\bibitem [{\citenamefont {Chen}\ \emph {et~al.}(2022)\citenamefont {Chen},
  \citenamefont {Reed}, \citenamefont {MacKellar}, \citenamefont {Downes},
  \citenamefont {Almuhawish}, \citenamefont {Jamieson}, \citenamefont {Adams},\
  and\ \citenamefont {Weatherill}}]{Chen2022}%
  \BibitemOpen
  \bibfield  {author} {\bibinfo {author} {\bibfnamefont {S.}~\bibnamefont
  {Chen}}, \bibinfo {author} {\bibfnamefont {D.~J.}\ \bibnamefont {Reed}},
  \bibinfo {author} {\bibfnamefont {A.~R.}\ \bibnamefont {MacKellar}}, \bibinfo
  {author} {\bibfnamefont {L.~A.}\ \bibnamefont {Downes}}, \bibinfo {author}
  {\bibfnamefont {N.~F.~A.}\ \bibnamefont {Almuhawish}}, \bibinfo {author}
  {\bibfnamefont {M.~J.}\ \bibnamefont {Jamieson}}, \bibinfo {author}
  {\bibfnamefont {C.~S.}\ \bibnamefont {Adams}}, \ and\ \bibinfo {author}
  {\bibfnamefont {K.~J.}\ \bibnamefont {Weatherill}},\ }\href
  {https://opg.optica.org/optica/fulltext.cfm?uri=optica-9-5-485&id=472380}
  {\bibfield  {journal} {\bibinfo  {journal} {Optica}\ }\textbf {\bibinfo
  {volume} {9}},\ \bibinfo {pages} {485} (\bibinfo {year} {2022})}\BibitemShut
  {NoStop}%
\bibitem [{\citenamefont {Griffith}\ \emph {et~al.}(2010)\citenamefont
  {Griffith}, \citenamefont {Knappe},\ and\ \citenamefont
  {Kitching}}]{Griffith2010}%
  \BibitemOpen
  \bibfield  {author} {\bibinfo {author} {\bibfnamefont {W.~C.}\ \bibnamefont
  {Griffith}}, \bibinfo {author} {\bibfnamefont {S.}~\bibnamefont {Knappe}}, \
  and\ \bibinfo {author} {\bibfnamefont {J.}~\bibnamefont {Kitching}},\ }\href
  {\doibase 10.1364/OE.18.027167} {\bibfield  {journal} {\bibinfo  {journal}
  {Opt. Express}\ }\textbf {\bibinfo {volume} {18}},\ \bibinfo {pages} {27167}
  (\bibinfo {year} {2010})}\BibitemShut {NoStop}%
\bibitem [{\citenamefont {Jim{\'e}nez-Mart{\'\i}nez}\ and\ \citenamefont
  {Knappe}()}]{JimenezBook2017}%
  \BibitemOpen
  \bibfield  {author} {\bibinfo {author} {\bibfnamefont {R.}~\bibnamefont
  {Jim{\'e}nez-Mart{\'\i}nez}}\ and\ \bibinfo {author} {\bibfnamefont
  {S.}~\bibnamefont {Knappe}},\ }\enquote {\bibinfo {title} {{Microfabricated
  Optically-Pumped Magnetometers}},}\ in\ \href
  {https://link.springer.com/chapter/10.1007/978-3-319-34070-8_17} {\emph
  {\bibinfo {booktitle} {{High Sensitivity Magnetometers}}}},\ \bibinfo
  {editor} {edited by\ \bibinfo {editor} {\bibfnamefont {A.}~\bibnamefont
  {Grosz}}, \bibinfo {editor} {\bibfnamefont {M.~J.}\ \bibnamefont
  {Haji-Sheikh}}, \ and\ \bibinfo {editor} {\bibfnamefont {S.~C.}\ \bibnamefont
  {Mukhopadhyay}}}\ (\bibinfo  {publisher} {Springer International
  Publishing},\ \bibinfo {address} {Cham})\ pp.\ \bibinfo {pages}
  {523--551}\BibitemShut {NoStop}%
\bibitem [{\citenamefont {Jiang}\ \emph {et~al.}(2023)\citenamefont {Jiang},
  \citenamefont {Zhai}, \citenamefont {Jiang}, \citenamefont {Wang},
  \citenamefont {Chen}, \citenamefont {Zhang}, \citenamefont {Wu},
  \citenamefont {Zhang}, \citenamefont {Zeng}, \citenamefont {Lin},
  \citenamefont {Wang},\ and\ \citenamefont {Jin}}]{Jiang2023}%
  \BibitemOpen
  \bibfield  {author} {\bibinfo {author} {\bibfnamefont {M.}~\bibnamefont
  {Jiang}}, \bibinfo {author} {\bibfnamefont {H.}~\bibnamefont {Zhai}},
  \bibinfo {author} {\bibfnamefont {C.}~\bibnamefont {Jiang}}, \bibinfo
  {author} {\bibfnamefont {J.}~\bibnamefont {Wang}}, \bibinfo {author}
  {\bibfnamefont {C.}~\bibnamefont {Chen}}, \bibinfo {author} {\bibfnamefont
  {Q.}~\bibnamefont {Zhang}}, \bibinfo {author} {\bibfnamefont
  {D.}~\bibnamefont {Wu}}, \bibinfo {author} {\bibfnamefont {B.}~\bibnamefont
  {Zhang}}, \bibinfo {author} {\bibfnamefont {Z.}~\bibnamefont {Zeng}},
  \bibinfo {author} {\bibfnamefont {J.}~\bibnamefont {Lin}}, \bibinfo {author}
  {\bibfnamefont {Y.}~\bibnamefont {Wang}}, \ and\ \bibinfo {author}
  {\bibfnamefont {P.}~\bibnamefont {Jin}},\ }\href {\doibase 10.1063/5.0149388}
  {\bibfield  {journal} {\bibinfo  {journal} {Applied Physics Letters}\
  }\textbf {\bibinfo {volume} {123}},\ \bibinfo {pages} {062406} (\bibinfo
  {year} {2023})}\BibitemShut {NoStop}%
\bibitem [{\citenamefont {Ledbetter}\ \emph {et~al.}(2008)\citenamefont
  {Ledbetter}, \citenamefont {Savukov}, \citenamefont {Budker}, \citenamefont
  {Shah}, \citenamefont {Knappe}, \citenamefont {Kitching}, \citenamefont
  {Michalak}, \citenamefont {Xu},\ and\ \citenamefont
  {Pines}}]{Ledbetter2008PNAS}%
  \BibitemOpen
  \bibfield  {author} {\bibinfo {author} {\bibfnamefont {M.~P.}\ \bibnamefont
  {Ledbetter}}, \bibinfo {author} {\bibfnamefont {I.~M.}\ \bibnamefont
  {Savukov}}, \bibinfo {author} {\bibfnamefont {D.}~\bibnamefont {Budker}},
  \bibinfo {author} {\bibfnamefont {V.}~\bibnamefont {Shah}}, \bibinfo {author}
  {\bibfnamefont {S.}~\bibnamefont {Knappe}}, \bibinfo {author} {\bibfnamefont
  {J.}~\bibnamefont {Kitching}}, \bibinfo {author} {\bibfnamefont {D.~J.}\
  \bibnamefont {Michalak}}, \bibinfo {author} {\bibfnamefont {S.}~\bibnamefont
  {Xu}}, \ and\ \bibinfo {author} {\bibfnamefont {A.}~\bibnamefont {Pines}},\
  }\href {http://dx.doi.org/10.1073/pnas.0711505105} {\bibfield  {journal}
  {\bibinfo  {journal} {Proceedings of the National Academy of Sciences}\
  }\textbf {\bibinfo {volume} {105}},\ \bibinfo {pages} {2286–2290} (\bibinfo
  {year} {2008})}\BibitemShut {NoStop}%
\bibitem [{\citenamefont {Kennedy}(2014)}]{kennedy2014PhDthesis}%
  \BibitemOpen
  \bibfield  {author} {\bibinfo {author} {\bibfnamefont {D.~J.}\ \bibnamefont
  {Kennedy}},\ }\emph {\bibinfo {title} {Progress in Microfluidic Nuclear
  Magnetic Resonance}},\ \href {https://escholarship.org/uc/item/0mh9q665}
  {Ph.D. thesis},\ \bibinfo  {school} {UC Berkeley} (\bibinfo {year}
  {2014})\BibitemShut {NoStop}%
\bibitem [{\citenamefont {Eills}\ \emph {et~al.}(2022)\citenamefont {Eills},
  \citenamefont {Hale},\ and\ \citenamefont {Utz}}]{Eills2022review}%
  \BibitemOpen
  \bibfield  {author} {\bibinfo {author} {\bibfnamefont {J.}~\bibnamefont
  {Eills}}, \bibinfo {author} {\bibfnamefont {W.}~\bibnamefont {Hale}}, \ and\
  \bibinfo {author} {\bibfnamefont {M.}~\bibnamefont {Utz}},\ }\href
  {https://www.sciencedirect.com/science/article/abs/pii/S0079656521000340}
  {\bibfield  {journal} {\bibinfo  {journal} {Progress in Nuclear Magnetic
  Resonance Spectroscopy}\ }\textbf {\bibinfo {volume} {128}},\ \bibinfo
  {pages} {44–69} (\bibinfo {year} {2022})}\BibitemShut {NoStop}%
\bibitem [{\citenamefont {Faivre}\ and\ \citenamefont
  {Sch{\"u}ler}(2008)}]{FaivreCR2008}%
  \BibitemOpen
  \bibfield  {author} {\bibinfo {author} {\bibfnamefont {D.}~\bibnamefont
  {Faivre}}\ and\ \bibinfo {author} {\bibfnamefont {D.}~\bibnamefont
  {Sch{\"u}ler}},\ }\href {\doibase 10.1021/cr078258w} {\bibfield  {journal}
  {\bibinfo  {journal} {Chemical Reviews}\ }\textbf {\bibinfo {volume} {108}},\
  \bibinfo {pages} {4875} (\bibinfo {year} {2008})}\BibitemShut {NoStop}%
\bibitem [{\citenamefont {Monteil}\ \emph {et~al.}(2019)\citenamefont
  {Monteil}, \citenamefont {Vallenet}, \citenamefont {Menguy}, \citenamefont
  {Benzerara}, \citenamefont {Barbe}, \citenamefont {Fouteau}, \citenamefont
  {Cruaud}, \citenamefont {Floriani}, \citenamefont {Viollier}, \citenamefont
  {Adryanczyk}, \citenamefont {Leonhardt}, \citenamefont {Faivre},
  \citenamefont {Pignol}, \citenamefont {L{\'o}pez-Garc{\'\i}a}, \citenamefont
  {Weld},\ and\ \citenamefont {Lefevre}}]{MonteilNM2019}%
  \BibitemOpen
  \bibfield  {author} {\bibinfo {author} {\bibfnamefont {C.~L.}\ \bibnamefont
  {Monteil}}, \bibinfo {author} {\bibfnamefont {D.}~\bibnamefont {Vallenet}},
  \bibinfo {author} {\bibfnamefont {N.}~\bibnamefont {Menguy}}, \bibinfo
  {author} {\bibfnamefont {K.}~\bibnamefont {Benzerara}}, \bibinfo {author}
  {\bibfnamefont {V.}~\bibnamefont {Barbe}}, \bibinfo {author} {\bibfnamefont
  {S.}~\bibnamefont {Fouteau}}, \bibinfo {author} {\bibfnamefont
  {C.}~\bibnamefont {Cruaud}}, \bibinfo {author} {\bibfnamefont
  {M.}~\bibnamefont {Floriani}}, \bibinfo {author} {\bibfnamefont
  {E.}~\bibnamefont {Viollier}}, \bibinfo {author} {\bibfnamefont
  {G.}~\bibnamefont {Adryanczyk}}, \bibinfo {author} {\bibfnamefont
  {N.}~\bibnamefont {Leonhardt}}, \bibinfo {author} {\bibfnamefont
  {D.}~\bibnamefont {Faivre}}, \bibinfo {author} {\bibfnamefont
  {D.}~\bibnamefont {Pignol}}, \bibinfo {author} {\bibfnamefont
  {P.}~\bibnamefont {L{\'o}pez-Garc{\'\i}a}}, \bibinfo {author} {\bibfnamefont
  {R.~J.}\ \bibnamefont {Weld}}, \ and\ \bibinfo {author} {\bibfnamefont
  {C.~T.}\ \bibnamefont {Lefevre}},\ }\href
  {https://www.nature.com/articles/s41564-019-0432-7} {\bibfield  {journal}
  {\bibinfo  {journal} {Nature Microbiology}\ }\textbf {\bibinfo {volume}
  {4}},\ \bibinfo {pages} {1088} (\bibinfo {year} {2019})}\BibitemShut
  {NoStop}%
\bibitem [{\citenamefont {Jofre}\ \emph {et~al.}(2023)\citenamefont {Jofre},
  \citenamefont {Romeu},\ and\ \citenamefont {Jofre-Roca}}]{Jofre2023}%
  \BibitemOpen
  \bibfield  {author} {\bibinfo {author} {\bibfnamefont {M.}~\bibnamefont
  {Jofre}}, \bibinfo {author} {\bibfnamefont {J.}~\bibnamefont {Romeu}}, \ and\
  \bibinfo {author} {\bibfnamefont {L.}~\bibnamefont {Jofre-Roca}},\ }\href
  {\doibase 10.1088/1367-2630/acb37a} {\bibfield  {journal} {\bibinfo
  {journal} {New Journal of Physics}\ }\textbf {\bibinfo {volume} {25}},\
  \bibinfo {pages} {013028} (\bibinfo {year} {2023})}\BibitemShut {NoStop}%
\bibitem [{\citenamefont {Xu}\ \emph {et~al.}(2006)\citenamefont {Xu},
  \citenamefont {Donaldson}, \citenamefont {Pines}, \citenamefont {Rochester},
  \citenamefont {Budker},\ and\ \citenamefont
  {Yashchuk}}]{Xu2006magneticparticles}%
  \BibitemOpen
  \bibfield  {author} {\bibinfo {author} {\bibfnamefont {S.}~\bibnamefont
  {Xu}}, \bibinfo {author} {\bibfnamefont {M.~H.}\ \bibnamefont {Donaldson}},
  \bibinfo {author} {\bibfnamefont {A.}~\bibnamefont {Pines}}, \bibinfo
  {author} {\bibfnamefont {S.~M.}\ \bibnamefont {Rochester}}, \bibinfo {author}
  {\bibfnamefont {D.}~\bibnamefont {Budker}}, \ and\ \bibinfo {author}
  {\bibfnamefont {V.~V.}\ \bibnamefont {Yashchuk}},\ }\href
  {http://dx.doi.org/10.1063/1.2400077} {\bibfield  {journal} {\bibinfo
  {journal} {Applied Physics Letters}\ }\textbf {\bibinfo {volume} {89}}
  (\bibinfo {year} {2006})}\BibitemShut {NoStop}%
\bibitem [{\citenamefont {Hoese}\ \emph {et~al.}(2021)\citenamefont {Hoese},
  \citenamefont {Koch}, \citenamefont {Bharadwaj}, \citenamefont {Lang},
  \citenamefont {Hadden}, \citenamefont {Yoshizaki}, \citenamefont
  {Giakoumaki}, \citenamefont {Ramponi}, \citenamefont {Jelezko}, \citenamefont
  {Eaton},\ and\ \citenamefont {Kubanek}}]{Hoese2021}%
  \BibitemOpen
  \bibfield  {author} {\bibinfo {author} {\bibfnamefont {M.}~\bibnamefont
  {Hoese}}, \bibinfo {author} {\bibfnamefont {M.~K.}\ \bibnamefont {Koch}},
  \bibinfo {author} {\bibfnamefont {V.}~\bibnamefont {Bharadwaj}}, \bibinfo
  {author} {\bibfnamefont {J.}~\bibnamefont {Lang}}, \bibinfo {author}
  {\bibfnamefont {J.~P.}\ \bibnamefont {Hadden}}, \bibinfo {author}
  {\bibfnamefont {R.}~\bibnamefont {Yoshizaki}}, \bibinfo {author}
  {\bibfnamefont {A.~N.}\ \bibnamefont {Giakoumaki}}, \bibinfo {author}
  {\bibfnamefont {R.}~\bibnamefont {Ramponi}}, \bibinfo {author} {\bibfnamefont
  {F.}~\bibnamefont {Jelezko}}, \bibinfo {author} {\bibfnamefont {S.~M.}\
  \bibnamefont {Eaton}}, \ and\ \bibinfo {author} {\bibfnamefont
  {A.}~\bibnamefont {Kubanek}},\ }\href
  {https://journals.aps.org/prapplied/abstract/10.1103/PhysRevApplied.15.054059}
  {\bibfield  {journal} {\bibinfo  {journal} {Phys. Rev. Appl.}\ }\textbf
  {\bibinfo {volume} {15}},\ \bibinfo {pages} {054059} (\bibinfo {year}
  {2021})}\BibitemShut {NoStop}%
\bibitem [{\citenamefont {Allert}\ \emph {et~al.}(2022)\citenamefont {Allert},
  \citenamefont {Bruckmaier}, \citenamefont {Neuling}, \citenamefont
  {Freire-Moschovitis}, \citenamefont {Liu}, \citenamefont {Schrepel},
  \citenamefont {Schätzle}, \citenamefont {Knittel}, \citenamefont {Hermans},\
  and\ \citenamefont {Bucher}}]{Allert2022}%
  \BibitemOpen
  \bibfield  {author} {\bibinfo {author} {\bibfnamefont {R.~D.}\ \bibnamefont
  {Allert}}, \bibinfo {author} {\bibfnamefont {F.}~\bibnamefont {Bruckmaier}},
  \bibinfo {author} {\bibfnamefont {N.~R.}\ \bibnamefont {Neuling}}, \bibinfo
  {author} {\bibfnamefont {F.~A.}\ \bibnamefont {Freire-Moschovitis}}, \bibinfo
  {author} {\bibfnamefont {K.~S.}\ \bibnamefont {Liu}}, \bibinfo {author}
  {\bibfnamefont {C.}~\bibnamefont {Schrepel}}, \bibinfo {author}
  {\bibfnamefont {P.}~\bibnamefont {Schätzle}}, \bibinfo {author}
  {\bibfnamefont {P.}~\bibnamefont {Knittel}}, \bibinfo {author} {\bibfnamefont
  {M.}~\bibnamefont {Hermans}}, \ and\ \bibinfo {author} {\bibfnamefont
  {D.~B.}\ \bibnamefont {Bucher}},\ }\href {\doibase 10.1039/D2LC00874B}
  {\bibfield  {journal} {\bibinfo  {journal} {Lab Chip}\ }\textbf {\bibinfo
  {volume} {22}},\ \bibinfo {pages} {4831} (\bibinfo {year}
  {2022})}\BibitemShut {NoStop}%
\bibitem [{\citenamefont {Mitchell}\ and\ \citenamefont
  {Palacios~Alvarez}(2020)}]{Mitchell2020}%
  \BibitemOpen
  \bibfield  {author} {\bibinfo {author} {\bibfnamefont {M.~W.}\ \bibnamefont
  {Mitchell}}\ and\ \bibinfo {author} {\bibfnamefont {S.}~\bibnamefont
  {Palacios~Alvarez}},\ }\href {\doibase 10.1103/RevModPhys.92.021001}
  {\bibfield  {journal} {\bibinfo  {journal} {Rev. Mod. Phys.}\ }\textbf
  {\bibinfo {volume} {92}},\ \bibinfo {pages} {021001} (\bibinfo {year}
  {2020})}\BibitemShut {NoStop}%
\bibitem [{\citenamefont {Bellini}\ \emph {et~al.}(2010)\citenamefont
  {Bellini}, \citenamefont {Vishnubhatla}, \citenamefont {Bragheri},
  \citenamefont {Ferrara}, \citenamefont {Minzioni}, \citenamefont {Ramponi},
  \citenamefont {Cristiani},\ and\ \citenamefont {Osellame}}]{Bellini2010}%
  \BibitemOpen
  \bibfield  {author} {\bibinfo {author} {\bibfnamefont {N.}~\bibnamefont
  {Bellini}}, \bibinfo {author} {\bibfnamefont {K.~C.}\ \bibnamefont
  {Vishnubhatla}}, \bibinfo {author} {\bibfnamefont {F.}~\bibnamefont
  {Bragheri}}, \bibinfo {author} {\bibfnamefont {L.}~\bibnamefont {Ferrara}},
  \bibinfo {author} {\bibfnamefont {P.}~\bibnamefont {Minzioni}}, \bibinfo
  {author} {\bibfnamefont {R.}~\bibnamefont {Ramponi}}, \bibinfo {author}
  {\bibfnamefont {I.}~\bibnamefont {Cristiani}}, \ and\ \bibinfo {author}
  {\bibfnamefont {R.}~\bibnamefont {Osellame}},\ }\href
  {https://opg.optica.org/oe/abstract.cfm?URI=oe-18-5-4679} {\bibfield
  {journal} {\bibinfo  {journal} {Opt. Express}\ }\textbf {\bibinfo {volume}
  {18}},\ \bibinfo {pages} {4679} (\bibinfo {year} {2010})}\BibitemShut
  {NoStop}%
\bibitem [{\citenamefont {Schaap}\ \emph {et~al.}(2012)\citenamefont {Schaap},
  \citenamefont {Rohrlack},\ and\ \citenamefont {Bellouard}}]{Schaap2012}%
  \BibitemOpen
  \bibfield  {author} {\bibinfo {author} {\bibfnamefont {A.}~\bibnamefont
  {Schaap}}, \bibinfo {author} {\bibfnamefont {T.}~\bibnamefont {Rohrlack}}, \
  and\ \bibinfo {author} {\bibfnamefont {Y.}~\bibnamefont {Bellouard}},\ }\href
  {\doibase 10.1039/C2LC21091F} {\bibfield  {journal} {\bibinfo  {journal} {Lab
  Chip}\ }\textbf {\bibinfo {volume} {12}},\ \bibinfo {pages} {1527} (\bibinfo
  {year} {2012})}\BibitemShut {NoStop}%
\bibitem [{\citenamefont {Memeo}\ \emph {et~al.}(2021)\citenamefont {Memeo},
  \citenamefont {Paiè}, \citenamefont {Sala}, \citenamefont {Castriotta},
  \citenamefont {Guercio}, \citenamefont {Vaccari}, \citenamefont {Osellame},
  \citenamefont {Bassi},\ and\ \citenamefont {Bragheri}}]{Memeo2021}%
  \BibitemOpen
  \bibfield  {author} {\bibinfo {author} {\bibfnamefont {R.}~\bibnamefont
  {Memeo}}, \bibinfo {author} {\bibfnamefont {P.}~\bibnamefont {Paiè}},
  \bibinfo {author} {\bibfnamefont {F.}~\bibnamefont {Sala}}, \bibinfo {author}
  {\bibfnamefont {M.}~\bibnamefont {Castriotta}}, \bibinfo {author}
  {\bibfnamefont {C.}~\bibnamefont {Guercio}}, \bibinfo {author} {\bibfnamefont
  {T.}~\bibnamefont {Vaccari}}, \bibinfo {author} {\bibfnamefont
  {R.}~\bibnamefont {Osellame}}, \bibinfo {author} {\bibfnamefont
  {A.}~\bibnamefont {Bassi}}, \ and\ \bibinfo {author} {\bibfnamefont
  {F.}~\bibnamefont {Bragheri}},\ }\href
  {https://onlinelibrary.wiley.com/doi/10.1002/jbio.202000396} {\bibfield
  {journal} {\bibinfo  {journal} {Journal of biophotonics}\ }\textbf {\bibinfo
  {volume} {14}},\ \bibinfo {pages} {e202000396} (\bibinfo {year}
  {2021})}\BibitemShut {NoStop}%
\bibitem [{\citenamefont {Lucivero}\ \emph {et~al.}(2022)\citenamefont
  {Lucivero}, \citenamefont {Zanoni}, \citenamefont {Corrielli}, \citenamefont
  {Osellame},\ and\ \citenamefont {Mitchell}}]{Lucivero2022}%
  \BibitemOpen
  \bibfield  {author} {\bibinfo {author} {\bibfnamefont {V.~G.}\ \bibnamefont
  {Lucivero}}, \bibinfo {author} {\bibfnamefont {A.}~\bibnamefont {Zanoni}},
  \bibinfo {author} {\bibfnamefont {G.}~\bibnamefont {Corrielli}}, \bibinfo
  {author} {\bibfnamefont {R.}~\bibnamefont {Osellame}}, \ and\ \bibinfo
  {author} {\bibfnamefont {M.~W.}\ \bibnamefont {Mitchell}},\ }\href
  {https://opg.optica.org/oe/fulltext.cfm?uri=oe-30-15-27149&id=478792}
  {\bibfield  {journal} {\bibinfo  {journal} {Opt. Express}\ }\textbf {\bibinfo
  {volume} {30}},\ \bibinfo {pages} {27149} (\bibinfo {year}
  {2022})}\BibitemShut {NoStop}%
\bibitem [{\citenamefont {Marcinkevi\v{c}ius}\ \emph
  {et~al.}(2001)\citenamefont {Marcinkevi\v{c}ius}, \citenamefont {Juodkazis},
  \citenamefont {Watanabe}, \citenamefont {Miwa}, \citenamefont {Matsuo},
  \citenamefont {Misawa},\ and\ \citenamefont {Nishii}}]{Marcinkevicius2001}%
  \BibitemOpen
  \bibfield  {author} {\bibinfo {author} {\bibfnamefont {A.}~\bibnamefont
  {Marcinkevi\v{c}ius}}, \bibinfo {author} {\bibfnamefont {S.}~\bibnamefont
  {Juodkazis}}, \bibinfo {author} {\bibfnamefont {M.}~\bibnamefont {Watanabe}},
  \bibinfo {author} {\bibfnamefont {M.}~\bibnamefont {Miwa}}, \bibinfo {author}
  {\bibfnamefont {S.}~\bibnamefont {Matsuo}}, \bibinfo {author} {\bibfnamefont
  {H.}~\bibnamefont {Misawa}}, \ and\ \bibinfo {author} {\bibfnamefont
  {J.}~\bibnamefont {Nishii}},\ }\href
  {https://opg.optica.org/ol/abstract.cfm?URI=ol-26-5-277} {\bibfield
  {journal} {\bibinfo  {journal} {Opt. Lett.}\ }\textbf {\bibinfo {volume}
  {26}},\ \bibinfo {pages} {277} (\bibinfo {year} {2001})}\BibitemShut
  {NoStop}%
\bibitem [{\citenamefont {Osellame}\ \emph {et~al.}(2011)\citenamefont
  {Osellame}, \citenamefont {Hoekstra}, \citenamefont {Cerullo},\ and\
  \citenamefont {Pollnau}}]{Osellame2011}%
  \BibitemOpen
  \bibfield  {author} {\bibinfo {author} {\bibfnamefont {R.}~\bibnamefont
  {Osellame}}, \bibinfo {author} {\bibfnamefont {H.}~\bibnamefont {Hoekstra}},
  \bibinfo {author} {\bibfnamefont {G.}~\bibnamefont {Cerullo}}, \ and\
  \bibinfo {author} {\bibfnamefont {M.}~\bibnamefont {Pollnau}},\ }\href
  {\doibase https://doi.org/10.1002/lpor.201000031} {\bibfield  {journal}
  {\bibinfo  {journal} {Laser \& Photonics Reviews}\ }\textbf {\bibinfo
  {volume} {5}},\ \bibinfo {pages} {442} (\bibinfo {year} {2011})}\BibitemShut
  {NoStop}%
\bibitem [{\citenamefont {Corrielli}\ \emph {et~al.}(2021)\citenamefont
  {Corrielli}, \citenamefont {Crespi},\ and\ \citenamefont
  {Osellame}}]{Corr21}%
  \BibitemOpen
  \bibfield  {author} {\bibinfo {author} {\bibfnamefont {G.}~\bibnamefont
  {Corrielli}}, \bibinfo {author} {\bibfnamefont {A.}~\bibnamefont {Crespi}}, \
  and\ \bibinfo {author} {\bibfnamefont {R.}~\bibnamefont {Osellame}},\ }\href
  {https://www.degruyter.com/document/doi/10.1515/nanoph-2021-0419/html}
  {\bibfield  {journal} {\bibinfo  {journal} {Nanophotonics}\ }\textbf
  {\bibinfo {volume} {10}},\ \bibinfo {pages} {3789} (\bibinfo {year}
  {2021})}\BibitemShut {NoStop}%
\bibitem [{\citenamefont {Pelucchi}\ \emph {et~al.}(2022)\citenamefont
  {Pelucchi}, \citenamefont {Fagas}, \citenamefont {Aharonovich}, \citenamefont
  {Englund}, \citenamefont {Figueroa}, \citenamefont {Gong}, \citenamefont
  {Hannes}, \citenamefont {Liu}, \citenamefont {Lu}, \citenamefont {Matsuda},
  \citenamefont {Pan}, \citenamefont {Schreck}, \citenamefont {Sciarrino},
  \citenamefont {Silberhorn}, \citenamefont {Wang},\ and\ \citenamefont
  {Jöns}}]{Pelucchi2022}%
  \BibitemOpen
  \bibfield  {author} {\bibinfo {author} {\bibfnamefont {E.}~\bibnamefont
  {Pelucchi}}, \bibinfo {author} {\bibfnamefont {G.}~\bibnamefont {Fagas}},
  \bibinfo {author} {\bibfnamefont {I.}~\bibnamefont {Aharonovich}}, \bibinfo
  {author} {\bibfnamefont {D.}~\bibnamefont {Englund}}, \bibinfo {author}
  {\bibfnamefont {E.}~\bibnamefont {Figueroa}}, \bibinfo {author}
  {\bibfnamefont {Q.}~\bibnamefont {Gong}}, \bibinfo {author} {\bibfnamefont
  {H.}~\bibnamefont {Hannes}}, \bibinfo {author} {\bibfnamefont
  {J.}~\bibnamefont {Liu}}, \bibinfo {author} {\bibfnamefont {C.-Y.}\
  \bibnamefont {Lu}}, \bibinfo {author} {\bibfnamefont {N.}~\bibnamefont
  {Matsuda}}, \bibinfo {author} {\bibfnamefont {J.-W.}\ \bibnamefont {Pan}},
  \bibinfo {author} {\bibfnamefont {F.}~\bibnamefont {Schreck}}, \bibinfo
  {author} {\bibfnamefont {F.}~\bibnamefont {Sciarrino}}, \bibinfo {author}
  {\bibfnamefont {C.}~\bibnamefont {Silberhorn}}, \bibinfo {author}
  {\bibfnamefont {J.}~\bibnamefont {Wang}}, \ and\ \bibinfo {author}
  {\bibfnamefont {K.~D.}\ \bibnamefont {Jöns}},\ }\href
  {https://doi.org/10.1038/s42254-021-00398-z} {\bibfield  {journal} {\bibinfo
  {journal} {Nature Reviews Physics}\ }\textbf {\bibinfo {volume} {4}},\
  \bibinfo {pages} {194} (\bibinfo {year} {2022})}\BibitemShut {NoStop}%
\bibitem [{Note1()}]{Note1}%
  \BibitemOpen
  \bibinfo {note} {We performed leak tests with different epoxy-sealed cells,
  observing no leaking up to \SI {150}{\celsius }.}\BibitemShut {Stop}%
\bibitem [{\citenamefont {Romalis}\ \emph {et~al.}(1997)\citenamefont
  {Romalis}, \citenamefont {Miron},\ and\ \citenamefont {Cates}}]{Romalis1997}%
  \BibitemOpen
  \bibfield  {author} {\bibinfo {author} {\bibfnamefont {M.~V.}\ \bibnamefont
  {Romalis}}, \bibinfo {author} {\bibfnamefont {E.}~\bibnamefont {Miron}}, \
  and\ \bibinfo {author} {\bibfnamefont {G.~D.}\ \bibnamefont {Cates}},\ }\href
  {\doibase 10.1103/PhysRevA.56.4569} {\bibfield  {journal} {\bibinfo
  {journal} {Phys. Rev. A}\ }\textbf {\bibinfo {volume} {56}},\ \bibinfo
  {pages} {4569} (\bibinfo {year} {1997})}\BibitemShut {NoStop}%
\bibitem [{\citenamefont {Seltzer}(2008)}]{SeltzerThesis}%
  \BibitemOpen
  \bibfield  {author} {\bibinfo {author} {\bibfnamefont {S.~J.}\ \bibnamefont
  {Seltzer}},\ }\emph {\bibinfo {title} {Developments in Alkali-Metal Atomic
  Magnetometry}},\ \href
  {https://www.proquest.com/openview/9f380bf68dfd5aeaf7da43847a5f65ec/1?pq-origsite=gscholar&cbl=18750}
  {Ph.D. thesis},\ \bibinfo  {school} {Princeton University} (\bibinfo {year}
  {2008})\BibitemShut {NoStop}%
\bibitem [{\citenamefont {Rosenberry}\ \emph {et~al.}(2007)\citenamefont
  {Rosenberry}, \citenamefont {Reyes}, \citenamefont {Tupa},\ and\
  \citenamefont {Gay}}]{Rosenberry2007}%
  \BibitemOpen
  \bibfield  {author} {\bibinfo {author} {\bibfnamefont {M.~A.}\ \bibnamefont
  {Rosenberry}}, \bibinfo {author} {\bibfnamefont {J.~P.}\ \bibnamefont
  {Reyes}}, \bibinfo {author} {\bibfnamefont {D.}~\bibnamefont {Tupa}}, \ and\
  \bibinfo {author} {\bibfnamefont {T.~J.}\ \bibnamefont {Gay}},\ }\href
  {\doibase 10.1103/PhysRevA.75.023401} {\bibfield  {journal} {\bibinfo
  {journal} {Phys. Rev. A}\ }\textbf {\bibinfo {volume} {75}},\ \bibinfo
  {pages} {023401} (\bibinfo {year} {2007})}\BibitemShut {NoStop}%
\bibitem [{\citenamefont {Scholtes}\ \emph {et~al.}(2014)\citenamefont
  {Scholtes}, \citenamefont {Woetzel}, \citenamefont {IJsselsteijn},
  \citenamefont {Schultze},\ and\ \citenamefont {Meyer}}]{Scholtes2014}%
  \BibitemOpen
  \bibfield  {author} {\bibinfo {author} {\bibfnamefont {T.}~\bibnamefont
  {Scholtes}}, \bibinfo {author} {\bibfnamefont {S.}~\bibnamefont {Woetzel}},
  \bibinfo {author} {\bibfnamefont {R.}~\bibnamefont {IJsselsteijn}}, \bibinfo
  {author} {\bibfnamefont {V.}~\bibnamefont {Schultze}}, \ and\ \bibinfo
  {author} {\bibfnamefont {H.-G.}\ \bibnamefont {Meyer}},\ }\href
  {https://doi.org/10.1007/s00340-014-5824-z} {\bibfield  {journal} {\bibinfo
  {journal} {Applied Physics B}\ }\textbf {\bibinfo {volume} {117}},\ \bibinfo
  {pages} {211} (\bibinfo {year} {2014})}\BibitemShut {NoStop}%
\bibitem [{\citenamefont {Dyer}\ \emph {et~al.}(2023)\citenamefont {Dyer},
  \citenamefont {McWilliam}, \citenamefont {Hunter}, \citenamefont {Ingleby},
  \citenamefont {Burt}, \citenamefont {Sharp}, \citenamefont {Mirando},
  \citenamefont {Griffin}, \citenamefont {Riis},\ and\ \citenamefont
  {McGilligan}}]{Dyer2023}%
  \BibitemOpen
  \bibfield  {author} {\bibinfo {author} {\bibfnamefont {S.}~\bibnamefont
  {Dyer}}, \bibinfo {author} {\bibfnamefont {A.}~\bibnamefont {McWilliam}},
  \bibinfo {author} {\bibfnamefont {D.}~\bibnamefont {Hunter}}, \bibinfo
  {author} {\bibfnamefont {S.}~\bibnamefont {Ingleby}}, \bibinfo {author}
  {\bibfnamefont {D.~P.}\ \bibnamefont {Burt}}, \bibinfo {author}
  {\bibfnamefont {O.}~\bibnamefont {Sharp}}, \bibinfo {author} {\bibfnamefont
  {F.}~\bibnamefont {Mirando}}, \bibinfo {author} {\bibfnamefont {P.~F.}\
  \bibnamefont {Griffin}}, \bibinfo {author} {\bibfnamefont {E.}~\bibnamefont
  {Riis}}, \ and\ \bibinfo {author} {\bibfnamefont {J.~P.}\ \bibnamefont
  {McGilligan}},\ }\href {\doibase 10.1063/5.0153881} {\bibfield  {journal}
  {\bibinfo  {journal} {Applied Physics Letters}\ }\textbf {\bibinfo {volume}
  {123}},\ \bibinfo {pages} {074001} (\bibinfo {year} {2023})}\BibitemShut
  {NoStop}%
\bibitem [{\citenamefont {Dupont-Roc}\ \emph {et~al.}(1969)\citenamefont
  {Dupont-Roc}, \citenamefont {Haroche},\ and\ \citenamefont
  {Cohen-Tannoudji}}]{Dupontroc1969}%
  \BibitemOpen
  \bibfield  {author} {\bibinfo {author} {\bibfnamefont {J.}~\bibnamefont
  {Dupont-Roc}}, \bibinfo {author} {\bibfnamefont {S.}~\bibnamefont {Haroche}},
  \ and\ \bibinfo {author} {\bibfnamefont {C.}~\bibnamefont
  {Cohen-Tannoudji}},\ }\href
  {https://www.sciencedirect.com/science/article/pii/0375960169904800?via%3Dihub}
  {\bibfield  {journal} {\bibinfo  {journal} {Physics Letters A}\ }\textbf
  {\bibinfo {volume} {28}},\ \bibinfo {pages} {638} (\bibinfo {year}
  {1969})}\BibitemShut {NoStop}%
\bibitem [{\citenamefont {Shah}\ \emph {et~al.}(2007)\citenamefont {Shah},
  \citenamefont {Knappe}, \citenamefont {Schwindt},\ and\ \citenamefont
  {Kitching}}]{Shah2007}%
  \BibitemOpen
  \bibfield  {author} {\bibinfo {author} {\bibfnamefont {V.}~\bibnamefont
  {Shah}}, \bibinfo {author} {\bibfnamefont {S.}~\bibnamefont {Knappe}},
  \bibinfo {author} {\bibfnamefont {P.~D.~D.}\ \bibnamefont {Schwindt}}, \ and\
  \bibinfo {author} {\bibfnamefont {J.}~\bibnamefont {Kitching}},\ }\href
  {https://doi.org/10.1038/nphoton.2007.201} {\bibfield  {journal} {\bibinfo
  {journal} {Nature Photonics}\ }\textbf {\bibinfo {volume} {1}},\ \bibinfo
  {pages} {649} (\bibinfo {year} {2007})}\BibitemShut {NoStop}%
\bibitem [{\citenamefont {Krzyzewski}\ \emph {et~al.}(2019)\citenamefont
  {Krzyzewski}, \citenamefont {Perry}, \citenamefont {Gerginov},\ and\
  \citenamefont {Knappe}}]{KrzyzewskiJAP2019}%
  \BibitemOpen
  \bibfield  {author} {\bibinfo {author} {\bibfnamefont {S.~P.}\ \bibnamefont
  {Krzyzewski}}, \bibinfo {author} {\bibfnamefont {A.~R.}\ \bibnamefont
  {Perry}}, \bibinfo {author} {\bibfnamefont {V.}~\bibnamefont {Gerginov}}, \
  and\ \bibinfo {author} {\bibfnamefont {S.}~\bibnamefont {Knappe}},\ }\href
  {https://doi.org/10.1063/1.5098088} {\bibfield  {journal} {\bibinfo
  {journal} {Journal of Applied Physics}\ }\textbf {\bibinfo {volume} {126}},\
  \bibinfo {pages} {044504} (\bibinfo {year} {2019})}\BibitemShut {NoStop}%
\bibitem [{\citenamefont {Tayler}\ \emph {et~al.}(2022)\citenamefont {Tayler},
  \citenamefont {Mouloudakis}, \citenamefont {Zetter}, \citenamefont {Hunter},
  \citenamefont {Lucivero}, \citenamefont {Bodenstedt}, \citenamefont
  {Parkkonen},\ and\ \citenamefont {Mitchell}}]{Tayler2022}%
  \BibitemOpen
  \bibfield  {author} {\bibinfo {author} {\bibfnamefont {M.~C.~D.}\
  \bibnamefont {Tayler}}, \bibinfo {author} {\bibfnamefont {K.}~\bibnamefont
  {Mouloudakis}}, \bibinfo {author} {\bibfnamefont {R.}~\bibnamefont {Zetter}},
  \bibinfo {author} {\bibfnamefont {D.}~\bibnamefont {Hunter}}, \bibinfo
  {author} {\bibfnamefont {V.~G.}\ \bibnamefont {Lucivero}}, \bibinfo {author}
  {\bibfnamefont {S.}~\bibnamefont {Bodenstedt}}, \bibinfo {author}
  {\bibfnamefont {L.}~\bibnamefont {Parkkonen}}, \ and\ \bibinfo {author}
  {\bibfnamefont {M.~W.}\ \bibnamefont {Mitchell}},\ }\href
  {https://link.aps.org/doi/10.1103/PhysRevApplied.18.014036} {\bibfield
  {journal} {\bibinfo  {journal} {Phys. Rev. Appl.}\ }\textbf {\bibinfo
  {volume} {18}},\ \bibinfo {pages} {014036} (\bibinfo {year}
  {2022})}\BibitemShut {NoStop}%
\bibitem [{\citenamefont {Zetter}\ \emph {et~al.}(2020)\citenamefont {Zetter},
  \citenamefont {J.~M\"{a}kinen}, \citenamefont {Iivanainen}, \citenamefont
  {Zevenhoven}, \citenamefont {Ilmoniemi},\ and\ \citenamefont
  {Parkkonen}}]{bfieldtools2}%
  \BibitemOpen
  \bibfield  {author} {\bibinfo {author} {\bibfnamefont {R.}~\bibnamefont
  {Zetter}}, \bibinfo {author} {\bibfnamefont {A.}~\bibnamefont
  {J.~M\"{a}kinen}}, \bibinfo {author} {\bibfnamefont {J.}~\bibnamefont
  {Iivanainen}}, \bibinfo {author} {\bibfnamefont {K.~C.~J.}\ \bibnamefont
  {Zevenhoven}}, \bibinfo {author} {\bibfnamefont {R.~J.}\ \bibnamefont
  {Ilmoniemi}}, \ and\ \bibinfo {author} {\bibfnamefont {L.}~\bibnamefont
  {Parkkonen}},\ }\href {http://dx.doi.org/10.1063/5.0016087} {\bibfield
  {journal} {\bibinfo  {journal} {Journal of Applied Physics}\ }\textbf
  {\bibinfo {volume} {128}} (\bibinfo {year} {2020})}\BibitemShut {NoStop}%
\bibitem [{\citenamefont {Castagna}\ and\ \citenamefont
  {Weis}(2011)}]{CastagnaPRA2011}%
  \BibitemOpen
  \bibfield  {author} {\bibinfo {author} {\bibfnamefont {N.}~\bibnamefont
  {Castagna}}\ and\ \bibinfo {author} {\bibfnamefont {A.}~\bibnamefont
  {Weis}},\ }\href {\doibase 10.1103/PhysRevA.84.053421} {\bibfield  {journal}
  {\bibinfo  {journal} {Phys. Rev. A}\ }\textbf {\bibinfo {volume} {84}},\
  \bibinfo {pages} {053421} (\bibinfo {year} {2011})}\BibitemShut {NoStop}%
\bibitem [{\citenamefont {Gerginov}\ \emph {et~al.}()\citenamefont {Gerginov},
  \citenamefont {Krzyzewski},\ and\ \citenamefont {Knappe}}]{Gerginov2017}%
  \BibitemOpen
  \bibfield  {author} {\bibinfo {author} {\bibfnamefont {V.}~\bibnamefont
  {Gerginov}}, \bibinfo {author} {\bibfnamefont {S.}~\bibnamefont
  {Krzyzewski}}, \ and\ \bibinfo {author} {\bibfnamefont {S.}~\bibnamefont
  {Knappe}},\ }\href
  {https://opg.optica.org/josab/abstract.cfm?URI=josab-34-7-1429} {\bibfield
  {journal} {\bibinfo  {journal} {J. Opt. Soc. Am. B}\ }\textbf {\bibinfo
  {volume} {34}},\ \bibinfo {pages} {1429}}\BibitemShut {NoStop}%
\bibitem [{\citenamefont {Jim\'{e}nez-Mart\'{i}nez}\ \emph
  {et~al.}()\citenamefont {Jim\'{e}nez-Mart\'{i}nez}, \citenamefont {Griffith},
  \citenamefont {Knappe}, \citenamefont {Kitching},\ and\ \citenamefont
  {Prouty}}]{JimenezMartinez2012}%
  \BibitemOpen
  \bibfield  {author} {\bibinfo {author} {\bibfnamefont {R.}~\bibnamefont
  {Jim\'{e}nez-Mart\'{i}nez}}, \bibinfo {author} {\bibfnamefont {W.~C.}\
  \bibnamefont {Griffith}}, \bibinfo {author} {\bibfnamefont {S.}~\bibnamefont
  {Knappe}}, \bibinfo {author} {\bibfnamefont {J.}~\bibnamefont {Kitching}}, \
  and\ \bibinfo {author} {\bibfnamefont {M.}~\bibnamefont {Prouty}},\ }\href
  {https://opg.optica.org/josab/abstract.cfm?URI=josab-29-12-3398} {\bibfield
  {journal} {\bibinfo  {journal} {J. Opt. Soc. Am. B}\ }\textbf {\bibinfo
  {volume} {29}},\ \bibinfo {pages} {3398}}\BibitemShut {NoStop}%
\bibitem [{\citenamefont {Troullinou}\ \emph {et~al.}(2023)\citenamefont
  {Troullinou}, \citenamefont {Lucivero},\ and\ \citenamefont
  {Mitchell}}]{Troullinou2023}%
  \BibitemOpen
  \bibfield  {author} {\bibinfo {author} {\bibfnamefont {C.}~\bibnamefont
  {Troullinou}}, \bibinfo {author} {\bibfnamefont {V.~G.}\ \bibnamefont
  {Lucivero}}, \ and\ \bibinfo {author} {\bibfnamefont {M.~W.}\ \bibnamefont
  {Mitchell}},\ }\href {\doibase 10.1103/PhysRevLett.131.133602} {\bibfield
  {journal} {\bibinfo  {journal} {Phys. Rev. Lett.}\ }\textbf {\bibinfo
  {volume} {131}},\ \bibinfo {pages} {133602} (\bibinfo {year}
  {2023})}\BibitemShut {NoStop}%
\bibitem [{\citenamefont {Ledbetter}\ \emph {et~al.}(2013)\citenamefont
  {Ledbetter}, \citenamefont {Savukov}, \citenamefont {Seltzer},\ and\
  \citenamefont {Budker}}]{Ledbetter2013}%
  \BibitemOpen
  \bibfield  {author} {\bibinfo {author} {\bibfnamefont {M.~P.}\ \bibnamefont
  {Ledbetter}}, \bibinfo {author} {\bibfnamefont {I.}~\bibnamefont {Savukov}},
  \bibinfo {author} {\bibfnamefont {S.~J.}\ \bibnamefont {Seltzer}}, \ and\
  \bibinfo {author} {\bibfnamefont {D.}~\bibnamefont {Budker}},\ }\enquote
  {\bibinfo {title} {Detection of nuclear magnetic resonance with atomic
  magnetometers},}\ in\ \href
  {https://www.cambridge.org/core/books/abs/optical-magnetometry/detection-of-nuclear-magnetic-resonance-with-atomic-magnetometers/EADBFEE71028DEEC039BD3C49AFD6CFD}
  {\emph {\bibinfo {booktitle} {Optical Magnetometry}}},\ \bibinfo {editor}
  {edited by\ \bibinfo {editor} {\bibfnamefont {D.}~\bibnamefont {Budker}}\
  and\ \bibinfo {editor} {\bibfnamefont {D.~F.}\ \bibnamefont
  {Jackson~Kimball}}}\ (\bibinfo  {publisher} {Cambridge University Press},\
  \bibinfo {year} {2013})\ p.\ \bibinfo {pages} {265–284}\BibitemShut
  {NoStop}%
\bibitem [{\citenamefont {Bajaj}\ \emph {et~al.}(2010)\citenamefont {Bajaj},
  \citenamefont {Paulsen}, \citenamefont {Harel},\ and\ \citenamefont
  {Pines}}]{Bajaj2010}%
  \BibitemOpen
  \bibfield  {author} {\bibinfo {author} {\bibfnamefont {V.~S.}\ \bibnamefont
  {Bajaj}}, \bibinfo {author} {\bibfnamefont {J.}~\bibnamefont {Paulsen}},
  \bibinfo {author} {\bibfnamefont {E.}~\bibnamefont {Harel}}, \ and\ \bibinfo
  {author} {\bibfnamefont {A.}~\bibnamefont {Pines}},\ }\href
  {https://www.science.org/doi/abs/10.1126/science.1192313} {\bibfield
  {journal} {\bibinfo  {journal} {Science}\ }\textbf {\bibinfo {volume}
  {330}},\ \bibinfo {pages} {1078} (\bibinfo {year} {2010})}\BibitemShut
  {NoStop}%
\bibitem [{\citenamefont {Jim{\'e}nez-Mart{\'\i}nez}\ \emph
  {et~al.}(2014)\citenamefont {Jim{\'e}nez-Mart{\'\i}nez}, \citenamefont
  {Kennedy}, \citenamefont {Rosenbluh}, \citenamefont {Donley}, \citenamefont
  {Knappe}, \citenamefont {Seltzer}, \citenamefont {Ring}, \citenamefont
  {Bajaj},\ and\ \citenamefont {Kitching}}]{Jimenez-MartinezNC2014}%
  \BibitemOpen
  \bibfield  {author} {\bibinfo {author} {\bibfnamefont {R.}~\bibnamefont
  {Jim{\'e}nez-Mart{\'\i}nez}}, \bibinfo {author} {\bibfnamefont {D.~J.}\
  \bibnamefont {Kennedy}}, \bibinfo {author} {\bibfnamefont {M.}~\bibnamefont
  {Rosenbluh}}, \bibinfo {author} {\bibfnamefont {E.~A.}\ \bibnamefont
  {Donley}}, \bibinfo {author} {\bibfnamefont {S.}~\bibnamefont {Knappe}},
  \bibinfo {author} {\bibfnamefont {S.~J.}\ \bibnamefont {Seltzer}}, \bibinfo
  {author} {\bibfnamefont {H.~L.}\ \bibnamefont {Ring}}, \bibinfo {author}
  {\bibfnamefont {V.~S.}\ \bibnamefont {Bajaj}}, \ and\ \bibinfo {author}
  {\bibfnamefont {J.}~\bibnamefont {Kitching}},\ }\href {\doibase
  10.1038/ncomms4908} {\bibfield  {journal} {\bibinfo  {journal} {Nature
  Communications}\ }\textbf {\bibinfo {volume} {5}},\ \bibinfo {pages} {3908}
  (\bibinfo {year} {2014})}\BibitemShut {NoStop}%
\bibitem [{\citenamefont {Theis}\ \emph {et~al.}(2011)\citenamefont {Theis},
  \citenamefont {Ganssle}, \citenamefont {Kervern}, \citenamefont {Knappe},
  \citenamefont {Kitching}, \citenamefont {Ledbetter}, \citenamefont {Budker},\
  and\ \citenamefont {Pines}}]{Theis2011}%
  \BibitemOpen
  \bibfield  {author} {\bibinfo {author} {\bibfnamefont {T.}~\bibnamefont
  {Theis}}, \bibinfo {author} {\bibfnamefont {P.}~\bibnamefont {Ganssle}},
  \bibinfo {author} {\bibfnamefont {G.}~\bibnamefont {Kervern}}, \bibinfo
  {author} {\bibfnamefont {S.}~\bibnamefont {Knappe}}, \bibinfo {author}
  {\bibfnamefont {J.}~\bibnamefont {Kitching}}, \bibinfo {author}
  {\bibfnamefont {M.~P.}\ \bibnamefont {Ledbetter}}, \bibinfo {author}
  {\bibfnamefont {D.}~\bibnamefont {Budker}}, \ and\ \bibinfo {author}
  {\bibfnamefont {A.}~\bibnamefont {Pines}},\ }\href
  {https://doi.org/10.1038/nphys1986} {\bibfield  {journal} {\bibinfo
  {journal} {Nature Physics}\ }\textbf {\bibinfo {volume} {7}},\ \bibinfo
  {pages} {571} (\bibinfo {year} {2011})}\BibitemShut {NoStop}%
\bibitem [{\citenamefont {Chuchkova}\ \emph {et~al.}(2023)\citenamefont
  {Chuchkova}, \citenamefont {Bodenstedt}, \citenamefont {Picazo-Frutos},
  \citenamefont {Eills}, \citenamefont {Tretiak}, \citenamefont {Hu},
  \citenamefont {Barskiy}, \citenamefont {de~Santis}, \citenamefont {Tayler},
  \citenamefont {Budker},\ and\ \citenamefont
  {Sheberstov}}]{ChuchkovaJPCL2023}%
  \BibitemOpen
  \bibfield  {author} {\bibinfo {author} {\bibfnamefont {L.}~\bibnamefont
  {Chuchkova}}, \bibinfo {author} {\bibfnamefont {S.}~\bibnamefont
  {Bodenstedt}}, \bibinfo {author} {\bibfnamefont {R.}~\bibnamefont
  {Picazo-Frutos}}, \bibinfo {author} {\bibfnamefont {J.}~\bibnamefont
  {Eills}}, \bibinfo {author} {\bibfnamefont {O.}~\bibnamefont {Tretiak}},
  \bibinfo {author} {\bibfnamefont {Y.}~\bibnamefont {Hu}}, \bibinfo {author}
  {\bibfnamefont {D.~A.}\ \bibnamefont {Barskiy}}, \bibinfo {author}
  {\bibfnamefont {J.}~\bibnamefont {de~Santis}}, \bibinfo {author}
  {\bibfnamefont {M.~C.~D.}\ \bibnamefont {Tayler}}, \bibinfo {author}
  {\bibfnamefont {D.}~\bibnamefont {Budker}}, \ and\ \bibinfo {author}
  {\bibfnamefont {K.~F.}\ \bibnamefont {Sheberstov}},\ }\href
  {https://doi.org/10.1021/acs.jpclett.3c01310} {\bibfield  {journal} {\bibinfo
   {journal} {The Journal of Physical Chemistry Letters}\ }\textbf {\bibinfo
  {volume} {14}},\ \bibinfo {pages} {6814} (\bibinfo {year}
  {2023})}\BibitemShut {NoStop}%
\bibitem [{\citenamefont {Mouloudakis}\ \emph {et~al.}(2023)\citenamefont
  {Mouloudakis}, \citenamefont {Bodenstedt}, \citenamefont {Azagra},
  \citenamefont {Mitchell}, \citenamefont {Marco-Rius},\ and\ \citenamefont
  {Tayler}}]{Mouloudakis2023}%
  \BibitemOpen
  \bibfield  {author} {\bibinfo {author} {\bibfnamefont {K.}~\bibnamefont
  {Mouloudakis}}, \bibinfo {author} {\bibfnamefont {S.}~\bibnamefont
  {Bodenstedt}}, \bibinfo {author} {\bibfnamefont {M.}~\bibnamefont {Azagra}},
  \bibinfo {author} {\bibfnamefont {M.~W.}\ \bibnamefont {Mitchell}}, \bibinfo
  {author} {\bibfnamefont {I.}~\bibnamefont {Marco-Rius}}, \ and\ \bibinfo
  {author} {\bibfnamefont {M.~C.~D.}\ \bibnamefont {Tayler}},\ }\href
  {https://doi.org/10.1021/acs.jpclett.2c03864} {\bibfield  {journal} {\bibinfo
   {journal} {The Journal of Physical Chemistry Letters}\ }\textbf {\bibinfo
  {volume} {14}},\ \bibinfo {pages} {1192} (\bibinfo {year}
  {2023})}\BibitemShut {NoStop}%
\bibitem [{\citenamefont {Khalil}\ \emph {et~al.}(2013)\citenamefont {Khalil},
  \citenamefont {Pichel}, \citenamefont {Zondervan}, \citenamefont {Abelmann},\
  and\ \citenamefont {Misra}}]{Khalil2013}%
  \BibitemOpen
  \bibfield  {author} {\bibinfo {author} {\bibfnamefont {I.~S.~M.}\
  \bibnamefont {Khalil}}, \bibinfo {author} {\bibfnamefont {M.~P.}\
  \bibnamefont {Pichel}}, \bibinfo {author} {\bibfnamefont {L.}~\bibnamefont
  {Zondervan}}, \bibinfo {author} {\bibfnamefont {L.}~\bibnamefont {Abelmann}},
  \ and\ \bibinfo {author} {\bibfnamefont {S.}~\bibnamefont {Misra}},\
  }\enquote {\bibinfo {title} {Characterization and control of biological
  microrobots},}\ in\ \href {\doibase 10.1007/978-3-319-00065-7_42} {\emph
  {\bibinfo {booktitle} {Experimental Robotics: The 13th International
  Symposium on Experimental Robotics}}},\ \bibinfo {editor} {edited by\
  \bibinfo {editor} {\bibfnamefont {J.~P.}\ \bibnamefont {Desai}}, \bibinfo
  {editor} {\bibfnamefont {G.}~\bibnamefont {Dudek}}, \bibinfo {editor}
  {\bibfnamefont {O.}~\bibnamefont {Khatib}}, \ and\ \bibinfo {editor}
  {\bibfnamefont {V.}~\bibnamefont {Kumar}}}\ (\bibinfo  {publisher} {Springer
  International Publishing},\ \bibinfo {address} {Heidelberg},\ \bibinfo {year}
  {2013})\ pp.\ \bibinfo {pages} {617--631}\BibitemShut {NoStop}%
\bibitem [{\citenamefont {Ceccarelli}\ \emph {et~al.}(2019)\citenamefont
  {Ceccarelli}, \citenamefont {Atzeni}, \citenamefont {Prencipe}, \citenamefont
  {Farinaro},\ and\ \citenamefont {Osellame}}]{Ceccarelli2019}%
  \BibitemOpen
  \bibfield  {author} {\bibinfo {author} {\bibfnamefont {F.}~\bibnamefont
  {Ceccarelli}}, \bibinfo {author} {\bibfnamefont {S.}~\bibnamefont {Atzeni}},
  \bibinfo {author} {\bibfnamefont {A.}~\bibnamefont {Prencipe}}, \bibinfo
  {author} {\bibfnamefont {R.}~\bibnamefont {Farinaro}}, \ and\ \bibinfo
  {author} {\bibfnamefont {R.}~\bibnamefont {Osellame}},\ }\href
  {https://opg.optica.org/jlt/abstract.cfm?URI=jlt-37-17-4275} {\bibfield
  {journal} {\bibinfo  {journal} {J. Lightwave Technol.}\ }\textbf {\bibinfo
  {volume} {37}},\ \bibinfo {pages} {4275} (\bibinfo {year}
  {2019})}\BibitemShut {NoStop}%
\bibitem [{\citenamefont {Ciriolo}\ \emph {et~al.}(2020)\citenamefont
  {Ciriolo}, \citenamefont {Vázquez}, \citenamefont {Tosa}, \citenamefont
  {Frezzotti}, \citenamefont {Crippa}, \citenamefont {Devetta}, \citenamefont
  {Faccialá}, \citenamefont {Frassetto}, \citenamefont {Poletto},
  \citenamefont {Pusala}, \citenamefont {Vozzi}, \citenamefont {Osellame},\
  and\ \citenamefont {Stagira}}]{Ciriolo2020}%
  \BibitemOpen
  \bibfield  {author} {\bibinfo {author} {\bibfnamefont {A.~G.}\ \bibnamefont
  {Ciriolo}}, \bibinfo {author} {\bibfnamefont {R.~M.}\ \bibnamefont
  {Vázquez}}, \bibinfo {author} {\bibfnamefont {V.}~\bibnamefont {Tosa}},
  \bibinfo {author} {\bibfnamefont {A.}~\bibnamefont {Frezzotti}}, \bibinfo
  {author} {\bibfnamefont {G.}~\bibnamefont {Crippa}}, \bibinfo {author}
  {\bibfnamefont {M.}~\bibnamefont {Devetta}}, \bibinfo {author} {\bibfnamefont
  {D.}~\bibnamefont {Faccialá}}, \bibinfo {author} {\bibfnamefont
  {F.}~\bibnamefont {Frassetto}}, \bibinfo {author} {\bibfnamefont
  {L.}~\bibnamefont {Poletto}}, \bibinfo {author} {\bibfnamefont
  {A.}~\bibnamefont {Pusala}}, \bibinfo {author} {\bibfnamefont
  {C.}~\bibnamefont {Vozzi}}, \bibinfo {author} {\bibfnamefont
  {R.}~\bibnamefont {Osellame}}, \ and\ \bibinfo {author} {\bibfnamefont
  {S.}~\bibnamefont {Stagira}},\ }\href {\doibase 10.1088/2515-7647/ab7d81}
  {\bibfield  {journal} {\bibinfo  {journal} {Journal of Physics: Photonics}\
  }\textbf {\bibinfo {volume} {2}},\ \bibinfo {pages} {024005} (\bibinfo {year}
  {2020})}\BibitemShut {NoStop}%
\bibitem [{\citenamefont {Vasilakis}\ \emph {et~al.}(2015)\citenamefont
  {Vasilakis}, \citenamefont {Shen}, \citenamefont {Jensen}, \citenamefont
  {Balabas}, \citenamefont {Salart}, \citenamefont {Chen},\ and\ \citenamefont
  {Polzik}}]{Vasilakis2015}%
  \BibitemOpen
  \bibfield  {author} {\bibinfo {author} {\bibfnamefont {G.}~\bibnamefont
  {Vasilakis}}, \bibinfo {author} {\bibfnamefont {H.}~\bibnamefont {Shen}},
  \bibinfo {author} {\bibfnamefont {K.}~\bibnamefont {Jensen}}, \bibinfo
  {author} {\bibfnamefont {M.}~\bibnamefont {Balabas}}, \bibinfo {author}
  {\bibfnamefont {D.}~\bibnamefont {Salart}}, \bibinfo {author} {\bibfnamefont
  {B.}~\bibnamefont {Chen}}, \ and\ \bibinfo {author} {\bibfnamefont {E.~S.}\
  \bibnamefont {Polzik}},\ }\href {\doibase 10.1038/nphys3280} {\bibfield
  {journal} {\bibinfo  {journal} {Nature Physics}\ }\textbf {\bibinfo {volume}
  {11}},\ \bibinfo {pages} {389} (\bibinfo {year} {2015})}\BibitemShut
  {NoStop}%
\bibitem [{\citenamefont {Ito}\ \emph {et~al.}(2016)\citenamefont {Ito},
  \citenamefont {Sato}, \citenamefont {Kamada},\ and\ \citenamefont
  {Kobayashi}}]{Ito2016}%
  \BibitemOpen
  \bibfield  {author} {\bibinfo {author} {\bibfnamefont {Y.}~\bibnamefont
  {Ito}}, \bibinfo {author} {\bibfnamefont {D.}~\bibnamefont {Sato}}, \bibinfo
  {author} {\bibfnamefont {K.}~\bibnamefont {Kamada}}, \ and\ \bibinfo {author}
  {\bibfnamefont {T.}~\bibnamefont {Kobayashi}},\ }\href
  {https://opg.optica.org/oe/abstract.cfm?URI=oe-24-14-15391} {\bibfield
  {journal} {\bibinfo  {journal} {Opt. Express}\ }\textbf {\bibinfo {volume}
  {24}},\ \bibinfo {pages} {15391} (\bibinfo {year} {2016})}\BibitemShut
  {NoStop}%
\bibitem [{\citenamefont {Kitching}(2018)}]{Kitching2018}%
  \BibitemOpen
  \bibfield  {author} {\bibinfo {author} {\bibfnamefont {J.}~\bibnamefont
  {Kitching}},\ }\href {https://doi.org/10.1063/1.5026238} {\bibfield
  {journal} {\bibinfo  {journal} {Applied Physics Reviews}\ }\textbf {\bibinfo
  {volume} {5}},\ \bibinfo {pages} {031302} (\bibinfo {year}
  {2018})}\BibitemShut {NoStop}%
\bibitem [{Note2()}]{Note2}%
  \BibitemOpen
  \bibinfo {note} {Obtained from the value $D_0=\SI {0.16}{\centi \meter
  \squared \per \second }$ measured in \cite {IshikawaPRA2000} at $T_0=\SI
  {60}{\celsius }$ using the scaling $D \propto (T(\SI {}{\kelvin })/T_0(\SI
  {}{\kelvin }))^{3/2}$ \cite {Lucivero2017}.}\BibitemShut {Stop}%
\bibitem [{Note3()}]{Note3}%
  \BibitemOpen
  \bibinfo {note} {For simplicity we neglect spin-exchange collisions between
  the two Rb isotopes and we use the spin-exchange broadening factor for
  $^{85}$Rb, obtained with nuclear spin $I=5/2$ in \cite
  {SeltzerThesis}}\BibitemShut {NoStop}%
\bibitem [{\citenamefont {Sheng}\ \emph {et~al.}(2013)\citenamefont {Sheng},
  \citenamefont {Li}, \citenamefont {Dural},\ and\ \citenamefont
  {Romalis}}]{Sheng2013}%
  \BibitemOpen
  \bibfield  {author} {\bibinfo {author} {\bibfnamefont {D.}~\bibnamefont
  {Sheng}}, \bibinfo {author} {\bibfnamefont {S.}~\bibnamefont {Li}}, \bibinfo
  {author} {\bibfnamefont {N.}~\bibnamefont {Dural}}, \ and\ \bibinfo {author}
  {\bibfnamefont {M.~V.}\ \bibnamefont {Romalis}},\ }\href {\doibase
  10.1103/PhysRevLett.110.160802} {\bibfield  {journal} {\bibinfo  {journal}
  {Phys. Rev. Lett.}\ }\textbf {\bibinfo {volume} {110}},\ \bibinfo {pages}
  {160802} (\bibinfo {year} {2013})}\BibitemShut {NoStop}%
\bibitem [{\citenamefont {Limes}\ \emph {et~al.}(2020)\citenamefont {Limes},
  \citenamefont {Foley}, \citenamefont {Kornack}, \citenamefont {Caliga},
  \citenamefont {McBride}, \citenamefont {Braun}, \citenamefont {Lee},
  \citenamefont {Lucivero},\ and\ \citenamefont {Romalis}}]{Limes2020}%
  \BibitemOpen
  \bibfield  {author} {\bibinfo {author} {\bibfnamefont {M.}~\bibnamefont
  {Limes}}, \bibinfo {author} {\bibfnamefont {E.}~\bibnamefont {Foley}},
  \bibinfo {author} {\bibfnamefont {T.}~\bibnamefont {Kornack}}, \bibinfo
  {author} {\bibfnamefont {S.}~\bibnamefont {Caliga}}, \bibinfo {author}
  {\bibfnamefont {S.}~\bibnamefont {McBride}}, \bibinfo {author} {\bibfnamefont
  {A.}~\bibnamefont {Braun}}, \bibinfo {author} {\bibfnamefont
  {W.}~\bibnamefont {Lee}}, \bibinfo {author} {\bibfnamefont {V.}~\bibnamefont
  {Lucivero}}, \ and\ \bibinfo {author} {\bibfnamefont {M.}~\bibnamefont
  {Romalis}},\ }\href
  {https://link.aps.org/doi/10.1103/PhysRevApplied.14.011002} {\bibfield
  {journal} {\bibinfo  {journal} {Phys. Rev. Appl.}\ }\textbf {\bibinfo
  {volume} {14}},\ \bibinfo {pages} {011002} (\bibinfo {year}
  {2020})}\BibitemShut {NoStop}%
\bibitem [{\citenamefont {Ishikawa}\ and\ \citenamefont
  {Yabuzaki}(2000)}]{IshikawaPRA2000}%
  \BibitemOpen
  \bibfield  {author} {\bibinfo {author} {\bibfnamefont {K.}~\bibnamefont
  {Ishikawa}}\ and\ \bibinfo {author} {\bibfnamefont {T.}~\bibnamefont
  {Yabuzaki}},\ }\href {\doibase 10.1103/PhysRevA.62.065401} {\bibfield
  {journal} {\bibinfo  {journal} {Phys. Rev. A}\ }\textbf {\bibinfo {volume}
  {62}},\ \bibinfo {pages} {065401} (\bibinfo {year} {2000})}\BibitemShut
  {NoStop}%
\bibitem [{\citenamefont {Lucivero}\ \emph {et~al.}(2017)\citenamefont
  {Lucivero}, \citenamefont {McDonough}, \citenamefont {Dural},\ and\
  \citenamefont {Romalis}}]{Lucivero2017}%
  \BibitemOpen
  \bibfield  {author} {\bibinfo {author} {\bibfnamefont {V.~G.}\ \bibnamefont
  {Lucivero}}, \bibinfo {author} {\bibfnamefont {N.~D.}\ \bibnamefont
  {McDonough}}, \bibinfo {author} {\bibfnamefont {N.}~\bibnamefont {Dural}}, \
  and\ \bibinfo {author} {\bibfnamefont {M.~V.}\ \bibnamefont {Romalis}},\
  }\href {\doibase 10.1103/PhysRevA.96.062702} {\bibfield  {journal} {\bibinfo
  {journal} {Phys. Rev. A}\ }\textbf {\bibinfo {volume} {96}},\ \bibinfo
  {pages} {062702} (\bibinfo {year} {2017})}\BibitemShut {NoStop}%
\end{thebibliography}%

\end{document}